\documentclass[11pt,a4paper]{article}
\usepackage{graphicx}
\usepackage{amssymb}
\usepackage{amsmath}
\usepackage{amsfonts}
\usepackage{dsfont}
\usepackage{mathtools}
\usepackage{array}
\usepackage{rotating}
\usepackage{bbold,amsfonts}
\allowdisplaybreaks

\usepackage[utf8]{inputenc}
\usepackage{bm}
\usepackage{xcolor}
\usepackage{float}
\usepackage{braket}
\usepackage{placeins}
\usepackage[height=8.8in,width=6.45in]{geometry}
\usepackage[font=small,labelfont=bf]{caption}
\usepackage[hidelinks]{hyperref}
\usepackage{booktabs,float,slashed}
\usepackage{standalone}
\usepackage{caption}
\usepackage{subcaption}
\usepackage{makecell}
\usepackage[maxbibnames=99,sorting=none,giveninits=true,backend=biber,style=numeric-comp,sortcites,doi=false,hyperref=true]{biblatex}
\renewbibmacro{in:}{}
\usepackage{hyperref}
\DeclareFieldFormat{doilink}{\iffieldundef{doi}{#1}{\href{https://doi.org/\thefield{doi}}{#1}}}
\DeclareFieldFormat[article,periodical]{volume}{\mkbibbold{#1}}
\DeclareFieldFormat[article,periodical]{journaltitle}{#1}
\DeclareFieldFormat[article,periodical]{pages}{#1}

\usepackage{xpatch}
\xpatchbibdriver{article}
  {\usebibmacro{journal+issuetitle}%
   \newunit
   \usebibmacro{byeditor+others}%
   \newunit
   \usebibmacro{note+pages}}
  {\printtext[doilink]{%
     \usebibmacro{journal+issuetitle}%
     \newunit
     \usebibmacro{byeditor+others}%
     \newunit
     \usebibmacro{note+pages}}}
  {}{}

\numberwithin{equation}{section}
\newcommand{\vev}[1]{{\left\langle #1 \right\rangle}}

\newcommand{\beq}{\begin{equation}}
\newcommand{\eeq}{\end{equation}}

\newcommand{\overbar}[1]{\mkern 1.5mu\overline{\mkern-1.5mu#1\mkern-1.5mu}\mkern 1.5mu}

\DeclareMathOperator{\tr}{tr}
\DeclareMathOperator{\diag}{diag}

\newcommand{\ii}{\mathrm{i}}

\makeatletter
\newcommand*{\letterdef@}{}
\newcommand*{\letterdef}[3]{%
	\def\letterdef@##1{\expandafter\newcommand\csname #1\endcsname{#2{##1}}}%
	\@tfor\@tempa :=#3\do{\expandafter\letterdef@\expandafter{\@tempa}}}
\makeatother
\letterdef{c#1} {\mathcal}{ABCDEFGHIJKLMNOPQRSTUVWXYZ} 
\letterdef{rm#1}{\mathrm} {dDeimM} 

\begin{document}
\begin{titlepage}
\vspace*{10mm}
\begin{center}
{\LARGE \bf 
Integrated correlators with a Wilson line \\[2mm]
in $\mathcal{N}=4$ SYM
}

\vspace*{15mm}

{\Large M. Bill\`o${}^{\,a,b}$, F. Galvagno${}^{\,c}$, M. Frau${}^{\,a,b}$, A. Lerda${}^{\,d,b}$}

\vspace*{8mm}
	
${}^a$ Universit\`a di Torino, Dipartimento di Fisica,\\
			Via P. Giuria 1, I-10125 Torino, Italy
			\vskip 0.3cm
			
${}^b$   I.N.F.N. - sezione di Torino,\\
			Via P. Giuria 1, I-10125 Torino, Italy 
			\vskip 0.3cm
   
${}^c$  Institut für Theoretische Physik, ETH Zürich,\\
Wolfgang-Pauli-Strasse 27, CH-8093 Zürich, Switzerland
			\vskip 0.3cm
   
${}^d$  Universit\`a del Piemonte Orientale,\\
			Dipartimento di Scienze e Innovazione Tecnologica\\
			Viale T. Michel 11, I-15121 Alessandria, Italy

\vskip 0.8cm
	{\small
		E-mail:
		\texttt{billo,frau,lerda@to.infn.it; francescogalvagn489@gmail.com}
	}
\vspace*{0.8cm}
\end{center}

\begin{abstract}

In the context of integrated correlators in $\cN=4$ SYM, we study the 2-point functions of local operators with a superconformal line defect. Starting from the mass-deformed $\cN=2^*$ theory in presence of a $\frac{1}{2}$-BPS Wilson line, we exploit the residual superconformal symmetry after the defect insertion, and show that the massive deformation corresponds to integrated insertions of the superconformal primaries belonging to the stress tensor multiplet with a specific integration measure which is explicitly derived after enforcing the superconformal Ward identities.
Finally, we show how the Wilson line integrated correlator can be computed by the $\cN=2^*$ Wilson loop vacuum expectation value on a 4-sphere in terms of a matrix model using supersymmetric localization. In particular, we reformulate previous matrix model computations by making use of recursion relations and Bessel kernels, providing a direct link with more general localization computations in $\cN=2$ theories.

\end{abstract}
\end{titlepage}
\setcounter{tocdepth}{2}
\tableofcontents

\section{Introduction}
\label{sec:intro}
A very interesting recent development in the study of the maximally supersymmetric $\cN=4$ Super Yang-Mills (SYM) theory in four dimensions concerns the correlators of scalar operators belonging to the stress-tensor multiplet. It is well-known that while the 2- and 3-point functions of such operators are fully protected by superconformal symmetry \cite{Lee:1998bxa}, their 4-point functions can be reduced to depending on a single function of the conformally invariant cross-ratios \cite{Dolan:2000ut,Nirschl:2004pa}. The new approach, initiated in \cite{Binder:2019jwn}, consists in integrating these 4-point functions with a certain measure prescribed by superconformal symmetry. These integrated correlators, which depend only on the rank of the gauge group and the gauge coupling, can be computed exactly using supersymmetric localization  \cite{Pestun:2007rz} by relating the stress-tensor multiplet insertions to the derivatives of the partition function of the $\cN=2^*$ theory, defined as a massive deformation of $\cN=4$ SYM that preserves $\cN=2$ supersymmetry.

Schematically, the two most studied integrated correlators read:
\begin{equation}
\begin{aligned}
 \int\!d^4x_1\dots d^4x_4
 ~ ~\mu(x_1,\dots, x_4) ~\big\langle
O_p(x_1)\, O_p(x_2)\,O_2(x_3) O_2(x_4)\big\rangle &= \partial_{\tau_p} \partial_{\bar{\tau_p}}\partial_{m^2} \log \cZ_{\cN=2^*}\Big|_{m=0}~, \\
\int\!d^4x_1\dots d^4x_4~ ~\mu^\prime(x_1,\dots, x_4) ~\big\langle O_2(x_1)\dots O_2(x_4)\big\rangle &= \partial_{m^4} \log \cZ_{\cN=2^*} \Big|_{m=0}~,
\end{aligned}
\label{integrated_corr}
\end{equation}
where $O_p$ is a $\frac{1}{2}$-BPS superconformal primary scalar operator with conformal dimension $p$ transforming in the $[0,p,0]$ representation of the $R$-symmetry gropus SU$(4)_R$ of the $\cN=4$ SYM. The case $p=2$ corresponds to the $\mathbf{20}^\prime$ scalar operator inside the stress-tensor multiplet. In the right-hand side of (\ref{integrated_corr}), the partition function $\cZ_{\cN=2^*}$ is computed with supersymmetric localization in terms of a matrix model and depends on the mass parameter $m$ (eventually set to 0) and on the couplings $\tau_p$ and $\overbar{\tau}_p$ associated to $O_p$ \cite{Gerchkovitz:2016gxx}. Notice that while the correlators in the left-hand side are computed in the flat Euclidean space $\mathbb{R}^4$, the partition function of the $\cN=2^*$ theory is defined on the 4-sphere $S^4$. It is precisely the conformal map between $S^4$ and $\mathbb R^4$, together with supersymmetry, that fixes the integration measures $\mu$ and $\mu^\prime$, which are the
crucial ingredients in the relations (\ref{integrated_corr}).
In particular, as we will recall below, each derivative with respect to $m$ determines the integrated insertion of a linear combination of different scalar operators, all belonging to the stress-tensor multiplet. Hence, after applying the constraints coming from superconformal Ward identities, the resulting correlators are all defined by a unique function of the conformal cross-ratios and can be integrated over a specific measure.  

The relations (\ref{integrated_corr}) were originally derived in \cite{Binder:2019jwn}, and further refined in \cite{Chester:2019jas,Chester:2020dja,Dorigoni:2021bvj,Dorigoni:2021guq}.
Many aspects of these integrated correlators have been subsequently explored, in particular by identifying their modular \cite{Chester:2020vyz,Collier:2022emf,Dorigoni:2022cua,Paul:2022piq} and weak-coupling \cite{Wen:2022oky} properties, by introducing general gauge groups \cite{Dorigoni:2022zcr,Dorigoni:2023}, by considering operator insertions with generic \cite{Brown:2023zbr} or large conformal dimensions \cite{Paul:2023rka,Brown:2023cpz,Brown:2023why,Caetano:2023zwe}, or by extending them to a pure $\cN=2$ setup \cite{Chester:2022sqb,Fiol:2023cml,Behan:2023fqq}.

Another very interesting class of integrated correlators is that of the 2-point functions in presence of a line defect, like for
example a $\frac{1}{2}$-BPS Wilson loop $W$. The idea comes from the supersymmetric localization results for the $\cN=2^*$ theory on $S^4$ obtained in \cite{Russo:2013kea,Russo:2013qaa,Buchel:2013id,Bobev:2013cja,Belitsky:2020hzs}, where not only the partition function but also the Wilson loop expectation value $\big\langle W\big\rangle_{\cN=2^*}$ was computed in terms of a matrix model. Following the same reasoning as before, the second mass derivative of $\log \big\langle W\big\rangle_{\cN=2^*}$ is expected to be associated to an integrated correlator of two $\mathbf{20^\prime}$ operators in presence of a Wilson loop, namely
\begin{equation}
\label{eq:1.2}
    \int\!d^4 x_1 d^4 x_2 ~~ \widehat{\mu}(x_1,x_2)~ \big\langle O_2(x_1) O_2(x_2)\big\rangle_W = \partial_{m^2} \log \big\langle W\big\rangle_{\cN=2^*}\Big|_{m=0}~.
\end{equation}
This relation has been conjectured in \cite{Pufu:2023vwo}, where the right-hand side has been studied in depth by computing the matrix model at strong coupling including several corrections beyond the planar level. Moreover, by applying a generic SL$(2,\mathbb{Z})$ transformation, the matrix model expectation value has been extended to generic $(p,q)$ dyonic extended particles, dual to extended $(p,q)$ strings in Anti-de Sitter.

In this paper we aim to deriving the conjectured formula \eqref{eq:1.2} concentrating our efforts on the left-hand side.  First of all, in Section~\ref{sec:review} we briefly review the main properties of the un-integrated correlator $\big\langle O_2(x_1) O_2(x_2)\big\rangle_W$ (referring to  \cite{Liendo:2016ymz,Barrat:2020vch,Barrat:2021yvp,Barrat:2022psm,Gimenez-Grau:2023fcy} for a more detailed analysis). In particular,
following the $R$-symmetry decomposition and the defect CFT prescriptions, we show that the dynamical part of $\big\langle O_2(x_1) O_2(x_2)\big\rangle_W$ can be written in terms of three functions of two cross-ratios, but only one of such functions, corresponding to the so-called 0-channel and denoted as $F_0$ in \cite{Barrat:2020vch,Barrat:2021yvp}, is sensitive to the mass deformation and is relevant for our purposes.

Next, after discussing in Section~\ref{sec:N2*} the main properties of the $\cN=2^*$ theory, in Section~\ref{sec:integrated} we focus on the explicit derivation of the integration measure 
$\widehat{\mu}(x_1,x_2)$, exploiting the stereographic projection from $S^4$ to $\mathbb R^4$ and the constraints arising from superconformal Ward identities. Compared to the case of the integrated 4-point functions, the presence of a $\frac{1}{2}$-BPS line defect partially breaks both supersymmetry and conformal invariance, leading to highly non-trivial relations among the correlators of operators belonging to the stress-tensor multiplet.

In Section~\ref{sec:matrix} we present the matrix model calculation
of the right-hand side of (\ref{eq:1.2}) based on the use recursion relations and Bessel kernels, along the lines recently developed in
\cite{Beccaria:2021hvt,Billo:2021rdb,Billo:2022xas,Billo:2022gmq,Billo:2022fnb,Billo:2022lrv} for the bulk 2- and 3-point functions in certain $\cN=2$ superconformal theories. Even though we recover already known results, we think that our derivation is worth being presented and useful
not only because it is simpler and more direct, but also because it is amenable for generalizations to $\cN=2$ SYM theories.

Finally, in Section~\ref{sec:conclusions} we present our conclusions and discuss some open problems. A lot of technical details are collected in four Appendices. In particular, Appendix~\ref{appendix:ward} contains an explicit derivation of the constraints on the 2-point functions in presence of a line defect
that are imposed by the symmetries of the model and of the Ward identities that they satisfy.

\section{Bulk correlators with a Wilson line: a quick review}
\label{sec:review}

We consider the $\cN=4$ SYM theory with gauge group SU($N$) defined in the flat Euclidean space $\mathbb{R}^4$, whose coordinates will be indicated by $x^\mu$ ($\mu=1,\cdots,4$). The fields of the theory are a vector $A_\mu$, six scalars $\phi^I$ ($I=1,\cdots 6)$, and four chiral fermions plus their anti-chiral counterparts. As is well-known, this field content guarantees superconformal invariance at the quantum level.

Since the early days of the AdS-CFT correspondence, a widely studied superconformal defect in this theory has been
the $\frac{1}{2}$-BPS Wilson line in the fundamental representation of SU$(N)$ \cite{Maldacena:1998im,Erickson:2000af,Drukker:2000rr}, defined by 
\begin{equation}
\label{defWLsusy}
W=\frac{1}{N}\tr \, \mathcal{P}
		\exp \bigg\{\sqrt{\frac{\lambda}{N}} \int \!d\tau \Big[\ii \,A_{\mu}(x)\,\dot{x}^{\mu}(\tau)
+ |\dot x(\tau)| \,\theta_I \,\phi^I(x)\Big] \bigg\}~.
\end{equation}
Here $\lambda$ is the 't Hooft coupling and $\theta_I$ are the components of a unit-normalized six-dimensional vector that parametrizes a direction in the space of the scalar fields $\phi^I$. Without any loss of generality, we choose the coordinate axes in such a way that the defect line is parametrized by $x^\mu(\tau) = (0,0,0,\tau)$. With this arrangement, then, the coordinates of a bulk point
$x$ are naturally divided into the three transverse ones $\vec{x}=(x^m)$ with $m=1,2,3$ and the longitudinal one $x^4$.

The insertion of $W$ breaks the global superconformal and $R$-symmetry algebra of $\cN=4$ SYM to a sub-algebra whose bosonic part is
\begin{equation}
\mathfrak{so}(1,2) \oplus \mathfrak{so}(3) \oplus \mathfrak{so}(5)_R~.
\label{breaking}
\end{equation}
The first two terms define the breaking pattern of the four-dimensional Euclidean conformal algebra 
$\mathfrak{so}(1,5)$ in presence of a line defect and account, respectively, for the conformal transformations along the line and for the rotations in the three-dimensional space transverse to the line. They represent also the isometries of AdS$_2\times S^2$ to which $\mathbb{R}^4$ with a line can be conformally mapped in such a way that the line becomes the boundary of AdS$_2$ 
(see for example \cite{Buchbinder:2012vr} for details). Finally, the last term in (\ref{breaking}) corresponds to the residual $R$-symmetry that remains after picking the scalar direction defined by the unit vector $\theta$ in the Wilson line connection.

The Wilson line also breaks half of the supersymmetries of the bulk theory. In fact it is invariant only under transformations in which the anti-chiral parameters are directly related to the chiral ones, see Appendices~\ref{appendix:spinor} and \ref{appendix:susy} for details. In other words the supersymmetry is effectively reduced to $\mathcal{N}=2$.

In presence of a conformal defect, bulk operators can have a non-vanishing 1-point function, whose space dependence is completely fixed by the residual conformal invariance \cite{Billo:2016cpy}. For example, the 1-point function of a scalar operator $O_\Delta$ with conformal dimension $\Delta$ in presence of $W$ is\,%
\footnote{If the Wilson line is normalized as in (\ref{defWLsusy}), one has $\langle W\rangle=1$.}
\begin{equation}
\label{eq:1pt_W}
\big\langle O_{\Delta} (x)\big\rangle_W \,\equiv\, \frac{\big\langle W\,O_\Delta(x) \big\rangle}{
\big\langle W \big\rangle} = \frac{a_{\Delta}}{|\vec{x}|^\Delta}~,
\end{equation}
where $|\vec{x}|$ represents the orthogonal distance between $x$ and the line, and the coefficient $a_{\Delta}$, which in general is a non-trivial function of $N$ and $\lambda$, is part of the conformal data.

The conformal kinematics, instead, does not fix the space-dependence of the bulk 2-point functions. In fact, in the case of two scalar operators $O_{\Delta_1}$ and $O_{\Delta_2}$ with conformal dimensions $\Delta_1$ and $\Delta_2$, the 2-point function can be written as
\begin{equation}
\label{eq:2pt_W}
    \big\langle O_{\Delta_1} (x_1)\,O_{\Delta_2} (x_2)\big\rangle_W = \frac{a_{\Delta_1\Delta_2} (\xi,\eta)}{|\vec{x}_1|^{\Delta_1}\,
    |\vec{x}_2|^{\Delta_2}}~,
\end{equation}
where the coefficient in the numerator depends on the two conformally invariant cross-ratios that can be constructed given two bulk points in presence of a line defect.
In the literature, various definitions of $\xi$ and $\eta$ are used. Here we find convenient to follow \cite{Buchbinder:2012vr} and take
\begin{equation}
    \xi=\frac{(x_{12}^4)^2+|\vec{x}_1|^2+
    |\vec{x}_2|^2}{2\,|\vec{x}_1|\,|\vec{x}_2|}~,\qquad
    \eta=\frac{\vec{x}_1\cdot \vec{x}_2}{|\vec{x}_1|\,|\vec{x}_2|}~,
  \label{eq:CrossRatios}
\end{equation}
where $x_{12}^4=(x_1^4-x_2^4)$,
which correspond to the geodesic distances between $x_1$ and $x_2$ in AdS$_2$ and $S^2$ respectively.

In our analysis we will often consider a particular class of bulk operators, namely the single-trace chiral primary operators that belong to the $\mathbf{20}^\prime$ symmetric traceless representation of the $\mathfrak{so}(6)_R$ $R$-symmetry algebra. 
These operators are part of the stress-tensor multiplet of $\cN=4$ SYM, are $\frac{1}{2}$-BPS and can be written as
\begin{equation}\label{eq:O20}
\cO(x,u) = u_I \,u_J\, \tr \phi^I(x) \phi^J(x)~,
\end{equation}
where $u_I$ are the six components of a vector $u$, such that $u^2=0$ to enforce the tracelessness condition.
The correlation functions of these operators in presence of a Wilson line have been widely studied from several different points of view \cite{Berenstein:1998ij,Semenoff:2001xp,Pestun:2002mr,Giombi:2009ds,Giombi:2012ep,Buchbinder:2012vr,Liendo:2016ymz,Barrat:2020vch,Barrat:2021yvp,Barrat:2022psm,Gimenez-Grau:2023fcy}. 
In particular, it has been shown that their 1-point function defines a coefficient, usually called $h_W$, which is related to the Bremsstrahlung function in $\cN=4$ SYM \cite{Correa:2012at,Correa:2012hh,Gromov:2012eu,Fiol:2012sg,Lewkowycz:2013laa,Giombi:2018hsx} \,%
\footnote{This relation has been generalized using supersymmetric localization to generic $\cN=2$ SCFTs \cite{Fiol:2015spa,Bianchi:2018zpb,Bianchi:2019dlw}.}. 

The correlator of two $\mathbf{20}^\prime$ operators has the generic form (\ref{eq:2pt_W}) with $\Delta_1=\Delta_2=2$. However, thanks to the residual superconformal symmetry, it is possible to introduce a further $R$-symmetry cross-ratio according to
\begin{align}
    \sigma = \frac{(u_1\cdot u_2)}{(u_1\cdot \theta)\,(u_2\cdot \theta)}~,
\end{align}
and write the 2-point function as
\begin{equation}\label{eq:2pt_O20_W}
    \big\langle \cO (x_1,u_1)\,\cO (x_2,u_2)\big\rangle_W = (u_1\cdot \theta)^2\,(u_2\cdot \theta)^2\, \frac{\cF(\xi,\eta,\sigma)}{|\vec{x}_1^\perp|^2\,|\vec{x}_2^\perp|^2}~.
\end{equation}
Here $\cF$ is a polynomial of order 2 in $\sigma$, namely
\begin{equation}\label{eq:Rsymmetry_channel}
    \cF(\xi,\eta,\sigma) = \sigma^2\,F_0(\xi,\eta)+\sigma\,F_1(\xi,\eta)+F_2(\xi,\eta) ~.
\end{equation}
The three functions $F_0$, $F_1$ and $F_2$ correspond to the three possible channels in which the polarization vectors $u_1$ and $u_2$
of the bulk operators contract, respectively, zero, one or two times with the $\theta$ vector of Wilson line.
Actually these three functions are not fully independent, since the following constraints have to be imposed \cite{Barrat:2020vch,Barrat:2021yvp}
\begin{equation}
    \Big( \partial_z + \frac{1}{2} \partial_\omega  \Big) \cF(\xi,\eta,\sigma) \bigg|_{z=\omega} =0~, \qquad  \Big( \partial_{\bar z} + \frac{1}{2} \partial_\omega  \Big) \cF(\xi,\eta,\sigma) \bigg|_{\bar z=\omega} =0~,
\end{equation}
where $(z,\bar z, \omega)$ are related to $(\xi,\eta, \sigma)$ according to
\begin{equation}
    \xi=\frac{1+ z\bar z}{2\sqrt{z\bar z}}~,\qquad
    \eta= \frac{z+\bar z}{2\sqrt{z\bar z}}~,\qquad
    \sigma= -\frac{(1-\omega)^2}{2\omega}~.
\end{equation}

As mentioned in the Introduction, our goal is to probe the $\cN=4$ SYM theory under a massive deformation that preserves half of its supersymmetries. To do this, we have to split the fields of $\cN=4$ SYM into a $\cN=2$ vector multiplet and a $\cN=2$ hyper-multiplet, and give a mass to the latter. Once this split is realized, the initial $R$-symmetry algebra is broken to
\begin{align}
    \mathfrak{su}(2)_F\oplus\mathfrak{su}(2)_R \oplus \mathfrak{u}(1)_R
\end{align}
where the first component is usually referred to as the flavor symmetry. Correspondingly, the $\mathbf{20}^\prime$ operators reorganize
as follows
\begin{equation}
    \mathbf{20}^\prime \rightarrow (\mathbf{1},\mathbf{1})_{\pm 2} \oplus(\mathbf{1},\mathbf{1})_{0}\oplus (\mathbf{2},\mathbf{2})_{\pm 1}\oplus (\mathbf{3},\mathbf{3})_{0}~.
    \label{split}
\end{equation}
Here, $(\mathbf{1},\mathbf{1})_{\pm 2}$ correspond to the chiral and anti-chiral dimension-2 operators made up with the scalars of the $\cN=2$ vector multiplet, while $(\mathbf{3},\mathbf{3})_{0}$ are three dimension-2 triplets of $\mathfrak{su}(2)_R$, aligned in the directions of $\mathfrak{su}(2)_F$, which are built only with the scalars of the hyper-multiplet. In the next section, following \cite{Binder:2019jwn}, we will consider a mass deformation of $\cN=4$ SYM that is driven by one of these triplets constructed with the hyper-multiplet scalars. On the other hand, in the Wilson line connection only the scalars of the $\cN=2$ vector multiplet can appear. Therefore, in the correlators of two bulk operators inducing the mass deformation in presence of a Wilson line, we necessarily have   
\begin{align}
    (u_1\cdot \theta)=(u_2\cdot\theta)=0~.
\end{align}
This means that only the 0-channel, with the corresponding function $F_0$, can be probed in this way.

One could also consider deformations associated to other components of the $\mathbf{20}^\prime$ decomposition \eqref{split} which correspond to integrated correlators\,%
\footnote{We expect that not all the deformations give rise to linearly independent integrated correlators, as shown in \cite{Chester:2020dja} for the integrated 4-point function.} probing different $R$-symmetry channels in the expansion \eqref{eq:Rsymmetry_channel}. For example the deformation associated to the marginal coupling $(\tau,\bar\tau)$ has been already studied in the literature \cite{Giombi:2009ds,Giombi:2012ep,Buchbinder:2012vr,Bonini:2014vta,Giombi:2018hsx,Giombi:2018qox} and realizes the integrated correlator $\partial_\tau \partial_{\bar\tau}\log \vev{W}$, which is associated to insertions of the chiral/anti-chiral operators $(\mathbf{1},\mathbf{1})_{\pm 2}$. Such insertions are localized by supersymmetry arguments \cite{Gerchkovitz:2016gxx}, and hence belong the topological sub-sector of $\cN=4$ SYM (see Appendix C of \cite{Pufu:2023vwo} for a more detailed analysis\,%
\footnote{In general the topological sub-sector contains chiral/anti-chiral primaries with higher conformal dimension $\Delta=p$, which realize a correlator $\partial_{\tau_p} \partial_{\bar\tau_p}\log \vev{W}$, see \cite{Beccaria:2020ykg} for explicit results for some values of $p$.}).
In this paper we concentrate on the mass deformation of $\cN=4$ SYM, giving rise to the integrated correlator \eqref{eq:1.2}.

We conclude this brief review by anticipating that in order to derive our results, we will need to consider not only bulk 2-point functions of scalar operators like those mentioned above, but also correlators involving spin-1 conserved currents, namely the currents that are part of the so-called $\mathcal{N}=2$ current multiplet. The expressions of these correlators follow from the general formalism of \cite{Lauria:2018klo} (see also \cite{Herzog:2020bqw}) and involve functions of the cross-ratios $\xi$ and $\eta$, which in our case are related to the function $F_0$ by differential relations that are a consequence of superconformal Ward identities.

\section{The mass-deformed \texorpdfstring{$\mathcal{N}=4$ SYM on a 4-sphere}{}}
\label{sec:N2*}
To describe the mass deformation of $\cN=4$ SYM in detail, we need first to introduce some further notation. We denote by $\varphi$ and $\overbar{\varphi}$ the two scalars of the $\cN=2$ vector multiplet
that remain massless (and that are used to construct the Wilson line connection). 
On the other hand, we
denote by $q$ and $\widetilde{q}$ the two complex scalars of the $\cN=2$ hyper-multiplet, and by
$\psi^a$ and $\overbar\psi^{\,a}$ (with $a=1,2$) their chiral and anti-chiral fermionic partners%
\footnote{We refer to Appendix~\ref{appendix:spinor} for our conventions on spinors and to Appendix~\ref{appendix:susy} for the supersymmetry transformations.}.
The scalars $q$ and $\widetilde{q}$ and their complex conjugates can be arranged in a $(2\times 2)$ matrix $Q^{ia}$ (with $i,a=1,2$) as follows
\begin{align}
    \big(Q^{ia}\big)=\begin{pmatrix}
        {q}^*&\widetilde{q}~\\
        -\widetilde{q}^{\,*}&q
    \end{pmatrix} ~.
    \label{Qin}
\end{align}
This notation allows to write the supersymmetry transformations in a compact form (see for example \cite{Festuccia:2020yff} and Appendix~\ref{appendix:susy}) and makes manifest the fact that
the hyper-multiplet scalars form a doublet with respect to both the $\mathfrak{su}(2)_R$
R-symmetry (labeled by $i$) and the $\mathfrak{su}(2)_F$ flavor
symmetry (labeled by $a$).

We then consider a deformation of $\cN=4$ SYM in which the hyper-multiplet becomes massive with a mass parameter $m$. The resulting theory has only $\mathcal{N}=2$ supersymmetry and is usually called $\mathcal{N}=2^*$ SYM. Since later we will use supersymmetric localization \cite{Pestun:2007rz}, we have put the theory on a compact space, which for simplicity we take to be a round 4-sphere $S^4$. The action we consider is therefore
\begin{equation}
    S_{\mathcal{N}=2^*}=S_{\mathcal{N}=4}+S_m
    \label{SN2star}
\end{equation}
where $S_{\mathcal{N}=4}$ is the standard $\mathcal{N}=4$ SYM action on a 4-sphere and \cite{Binder:2019jwn,Chester:2020dja,Pufu:2023vwo}\,%
\footnote{Notice that the action of the $\cN=2^*$ theory also contains a term proportional to $m$ and cubic in the scalar fields. However, such a term is proportional to the Yang-Mills coupling (see {\emph{e.g.}} \cite{Belitsky:2020hzs}) and thus contributes to terms of higher order. On the contrary the terms in (\ref{Sm}) which are the ones considered in \cite{Binder:2019jwn,Chester:2020dja,Pufu:2023vwo} are present also in the free theory and are sometimes called ``proper'' mass terms.}
\begin{align}
    S_m=\int\!d^4x\,\sqrt{g(x)}\,\bigg[
    m\Big(\frac{\ii}{R}\,J(x)+K(x)\Big)+m^2\,L(x)\bigg]~.
    \label{Sm}
\end{align}
Here $R$ is the radius of $S^4$ (which will be set to 1 from now on), $g(x)$ is the determinant of the metric of $S^4$, and 
    \begin{align}
    J(x)&=\tr q(x)q(x)+\tr \widetilde{q}(x)\,\widetilde{q}(x)+\tr q^*(x)q^*(x)+\tr \widetilde{q}^{\,*}(x)\,\widetilde{q}^{\,*}(x)
    =\tr Q^{ia}(x)Q^{ia}(x)~,
    \label{J}\\[2mm]
    K(x)&=-\ii \tr \psi^{a}(x)\psi^a(x) -\ii\,\tr \overbar{\psi}^a(x)\overbar{\psi}^{\,a}(x)
     \label{K}
     \end{align}
where the sum over repeated indices is understood.
The $m^2$-part of (\ref{Sm}) corresponds to the mass term for the bosons $q$ and $\widetilde{q}$ in flat space but, as argued in \cite{Binder:2019jwn,Chester:2020dja,Dorigoni:2021guq,Pufu:2023vwo}, it does not play
any significant role in the following analysis and can be effectively ignored\,%
\footnote{More precisely, in \cite{Binder:2019jwn,Chester:2020dja,Dorigoni:2021guq} it is argued that the contributions arising from the $m^2$-term cancel against the boundary contributions that are generated by integrations by parts. Thus, one can in practice ignore the $m^2$ terms and freely perform integrations by parts. \label{footnote_boundary}}. Thus, we may take \cite{Binder:2019jwn,Chester:2020dja,Pufu:2023vwo}
\begin{equation}
    S_m=m\int\!d^4x\,\sqrt{g(x)}\,
    \big(\ii\,J(x)+K(x)\big)~.
    \label{Sm1}
\end{equation}

The quadratic operators $J(x)$ and $K(x)$ are conformal primaries with dimensions 2 and 3 respectively,
and can be written in terms of the so-called $\mathcal{N}=2$ current multiplet which is a sub-multiplet of the $\mathcal{N}=4$ stress-tensor multiplet (see for example
\cite{Dolan:2001tt}). The field content of a current multiplet consists of
three scalars $\Phi^{ij}=\Phi^{ji}$ (with $i,j=1,2$) such that $(\Phi^{ij})^*=\epsilon_{ik}\,\epsilon_{j\ell}\,\Phi^{k\ell}$, two chiral fermions $X_i$ and two anti-chiral fermions $\overbar{X}_i$, two real scalars $P$ and $\overbar{P}$, and one conserved current $j_\mu$.
The supersymmetry transformations of these fields are provided in Appendix \ref{appendix:susy} where we also show that this multiplet can be realized in terms of the hyper-multiplet as follows
\begin{subequations}
    \begin{align}
    \Phi^{ij}&= \tr Q^{ia}\,Q^{ja}~,\label{Phiij}\\[1mm]
    X_i&=2\sqrt{2}\,\tr Q^{ja}\,\psi^a\,\epsilon_{ji}~,\quad
    \overbar{X}_i=2\sqrt{2}\,\tr Q^{ja}\,\overbar{\psi}^a\,\epsilon_{ji}~,\label{XXbar}\\[1mm]
    P&=2\,\tr \psi^a\,\psi^a~,\quad \overbar{P}=2\,\tr \overbar{\psi}^a\,\overbar{\psi}^a~,\label{PPbar}\\[1mm]
    j_\mu&=2\,\ii\,\tr Q^{ia}\,\partial_\mu Q^{ja}\,\epsilon_{ij}-2\,\tr {\psi}^{a}\sigma_\mu \overbar{\psi}^{\,a}~.
   \label{jmu}
\end{align}
 \label{flavorhyper}%
\end{subequations}
Notice that in these expressions the flavor indices $a$ of the constituent fields are summed over, meaning that the multiplet (\ref{flavorhyper}) is aligned along
a specific Cartan direction of $\mathfrak{su}(2)_F$. Furthermore, with this explicit realization one can check that the dimension-2 operators $\Phi^{ij}$ in (\ref{Phiij}) are the $\mathbf{20}^\prime$ operators that belong to one of the triplets in the representation $(\mathbf{3},\mathbf{3})_0$ of the residual $R$-symmetry algebra, as mentioned in Section~\ref{sec:review}.

Finally, comparing (\ref{J}) and (\ref{K}) with (\ref{Phiij}) and (\ref{PPbar}), we easily see that 
\begin{equation}
    J(x)=\Phi^{11}(x)+\Phi^{22}(x)~,\quad K(x)=-\frac{\ii}{2}\,P(x)-\frac{\ii}{2}\,\overbar{P}(x)~.
\label{JPhiKP}
\end{equation}
Thus, we can conclude that the mass deformation leading to $\cN=2^*$ SYM on a 4-sphere is driven by a particular combinations of the scalar components of a current multiplet along a Cartan direction of the residual flavor symmetry (see also Section 3 of \cite{Binder:2019jwn}
for further details). 
As a final remark, we note that since the massive deformation breaks half of the supersymmetry, all constraints that we will derive only require $\cN=2$ supersymmetry. However, when we restore $m=0$ at the end, these constraints bear some implications also for the function $F_0$ in \eqref{eq:Rsymmetry_channel} which is an observable of the $\cN=4$ theory in presence of a line defect.

\section{Integrated correlators with a Wilson line}
\label{sec:integrated}
We now consider a $\frac{1}{2}$-BPS line defect of the $\cN=2^*$ SYM theory on $S^4$ described by a circular
Wilson loop $W_\mathcal{C}$ in the fundamental representation of SU($N$). Its explicit expression is
\begin{equation}
\label{defWCsusy}
W_{\mathcal{C}}=\frac{1}{N}\tr \, \mathcal{P}
		\exp \bigg\{\sqrt{\frac{\lambda}{N}} \oint_{\cC} d\tau \Big[\ii \,A_{\mu}(x)\,\dot{x}^{\mu}(\tau)
+\frac{\varphi(x)+\overbar{\varphi}(x)}{\sqrt{2}}\Big] \bigg\}~.
\end{equation}
where $\mathcal{C}$ is a great circle of $S^4$ and, as before, $\lambda$ is the 't Hooft coupling.
As anticipated, the scalars that appear in the Wilson loop connections are those of the $\cN=2$ vector multiplet. In particular, the combination $\frac{\varphi+\overbar{\varphi}}{\sqrt{2}}$ simply corresponds to the scalar $\phi^1$ in the notation of Section~\ref{sec:review}, implying that the $\theta$-vector of $W_\cC$ is non-vanishing only in the first direction ({\it{i.e.}} $\theta^I=0$ for $I\not=1$). This means that there is no direct (or tree-level) coupling among the Wilson loop and the operators $J$ and $K$ defined in (\ref{J}) and (\ref{K}) which induce the massive deformation, and that all interactions among them are mediated by the $\cN=4$ SYM action.

The vacuum expectation value of $W_{\mathcal{C}}$ in the massive theory is
\begin{equation}
    \big\langle\!\!\big\langle W_{\mathcal{C}}\big\rangle\!\!\big\rangle_{\mathcal{N}=2^*} :=\,
    \frac{\displaystyle{\int \mathcal{D}[\mathrm{fields}]\,\,W_{\mathcal{C}}\,
    \rme^{-S_{\mathcal{N}=4}-S_m}}}{\displaystyle{\int \mathcal{D}[\mathrm{fields}]\,\,
    \rme^{-S_{\mathcal{N}=4}-S_m}}}
    \label{vevW}
\end{equation}
where $S_m$ is given in (\ref{Sm1}).
Expanding for small $m$, we have
\begin{align}
 \big\langle\!\!\big\langle W_{\mathcal{C}}\big\rangle\!\!\big\rangle_{\mathcal{N}=2^*} &=\big\langle\!\!\big\langle W_{\mathcal{C}}\big\rangle\!\!\big\rangle \bigg[1-
 \frac{\big\langle\!\!\big\langle
    W_\mathcal{C}\,S_m\big\rangle\!\!\big\rangle}{\big\langle\!\!\big\langle W_{\mathcal{C}}\big\rangle\!\!\big\rangle}+\frac{1}{2} \frac{\big\langle\!\!\big\langle
    W_\mathcal{C}\,S_m^2\big\rangle\!\!\big\rangle-\big\langle\!\!\big\langle
    W_\mathcal{C}\big\rangle\!\!\big\rangle\,\big\langle\!\!\big\langle S_m^2\big\rangle\!\!\big\rangle}{\big\langle\!\!\big\langle W_{\mathcal{C}}\big\rangle\!\!\big\rangle}+\ldots\bigg] ~.
    \label{expm}
\end{align}
Here we have used the symbol $\langle\!\langle~\rangle\!\rangle$ to denote the expectation values in the $\mathcal{N}=4$ SYM theory on $S^4$ to distinguish them from those in flat space that were simply indicated as $\langle~\rangle$. Moreover, in writing the expansion 
(\ref{expm}) we have exploited the fact that 
$\langle\!\langle S_m\rangle\!\rangle=0$. Notice that the contribution proportional to $m$, corresponding to the second term in the square bracket of (\ref{expm}), actually 
vanishes; indeed, since the Wilson loop $W_\mathcal{C}$ and $S_m$ are made up of fields belonging to multiplets
with different $R$-symmetry properties, the vacuum expectation value $\langle\!\langle W_{\mathcal{C}}\, S_m\rangle\!\rangle$ factorizes and thus vanishes:
$\langle\!\langle W_{\mathcal{C}}\, S_m\rangle\!\rangle=\langle\!\langle W_{\mathcal{C}}\rangle\!\rangle\,\langle\!\langle S_m\rangle\!\rangle=0$. Therefore,
$\partial_m \big\langle\!\!\big\langle W_{\mathcal{C}}\big\rangle\!\!\big\rangle_{\mathcal{N}=2^*} \big|_{m=0}=0$ and the first non-vanishing mass-correction is proportional to $m^2$. This contribution is captured by
\begin{align}
    \mathcal{I}= 
   \partial_m^2 \log \big\langle\!\!\big\langle W_{\mathcal{C}}\big\rangle\!\!\big\rangle_{\mathcal{N}=2^*} \Big|_{m=0}=\frac{\partial_m^2 \big\langle\!\!\big\langle W_{\mathcal{C}}\big\rangle\!\!\big\rangle_{\mathcal{N}=2^*} \Big|_{m=0}}{\big\langle\!\!\big\langle W_\mathcal{C}\big\rangle\!\!\big\rangle }\label{Iis}~,
\end{align}
which will be the main object of our analysis in the rest of this paper.

From (\ref{expm}) and the expression of $S_m$ given in (\ref{Sm1}), it follows that
\begin{equation}
\begin{aligned}
    \mathcal{I}
    &=-\int\!d^4x_1\,\sqrt{g(x_1)}
    \int\!d^4x_2\,\sqrt{g(x_2)}\,\,\frac{\big\langle\!\!\big\langle
    W_\mathcal{C}\,J(x_1)\,J(x_2)\big\rangle\!\!\big\rangle^{c}}{\big\langle\!\!\big\langle W_\mathcal{C} \big\rangle\!\!\big\rangle}\\[2mm]
    &\quad+\int\!d^4x_1\,\sqrt{g(x_1)}
    \int\!d^4x_2\,\sqrt{g(x_2)}\,\,\frac{\big\langle\!\!\big\langle
    W_\mathcal{C}\,K(x_1)\,K(x_2)\big\rangle\!\!\big\rangle^{c}}{\big\langle\!\!\big\langle W_\mathcal{C} \big\rangle\!\!\big\rangle}
    \label{Iis2}
\end{aligned}
\end{equation}
where the superscript $c$ stands for connected\,%
\footnote{We recall that, according to the observation in footnote \ref{footnote_boundary}, we have omitted the contribution of the 1-point function of the operator $L(x)$ appearing in the mass deformation (\ref{Sm}).}. 
Using a stereographic projection, one can map the correlators on $S^4$ to correlators in $\mathbb{R}^4$ and rewrite (\ref{Iis2}) as a combination of integrated bulk correlators in flat space, similar to those discussed in Section~\ref{sec:review}. In particular, adopting the stereographic projection described in Appendix~\ref{appendix:projection}, the
circular Wilson loop $W_\mathcal{C}$ in (\ref{defWCsusy}) can be mapped to a straight Wilson line $W$ along the $x^4$-direction in $\mathbb{R}^4$ as in (\ref{defWLsusy}) with $\theta=(1,0,0,0,0,0)$.
Furthermore, under this stereographic projection the correlator on $S^4$ of two bulk operators $\mathcal{O}$ of conformal dimension $\Delta$
is mapped to the corresponding correlator in $\mathbb{R}^4$ according to
\begin{equation}
    \big\langle\!\!\big\langle
    W_\mathcal{C}\,\mathcal{O}(x_1)\,\mathcal{O}(x_2)\big\rangle\!\!\big\rangle
    =\Big(\frac{1+x_1^2}{2}\Big)^\Delta\,\Big(\frac{1+x_2^2}{2}\Big)^\Delta\,
    \big\langle
    W\,\mathcal{O}(x_1)\,\mathcal{O}(x_2)\big\rangle
~.
\end{equation}
Using this formula for $\Delta=2$ and $\Delta=3$ in the two terms of (\ref{Iis2}), and exploiting also the explicit
expression of the determinant of the metric of $S^4$ in this parametrization, we obtain
\begin{equation}
\begin{aligned}
   \mathcal{I}&=-\int\!d^4x_1\!\int\!d^4x_2\,\frac{1}{\displaystyle{\Big(\frac{1+x_1^2}{2}\Big)^2\Big(\frac{1+x_2^2}{2}\Big)^2}}
    \,\,\big\langle
    J(x_1)\,J(x_2)\big\rangle_{W}\\[2mm]
    &\quad+\int\!d^4x_1\!\int\!d^4x_2\,\frac{1}{\displaystyle{\Big(\frac{1+x_1^2}{2}\Big)
    \Big(\frac{1+x_2^2}{2}\Big)}}
    \,\,\big\langle
    K(x_1)\,K(x_2)\big\rangle_{W}
    \label{Iis3}
\end{aligned}
\end{equation}
where we have used the same notation of Section~\ref{sec:review} for the bulk 2-point functions in presence of the Wilson line, dropping the superscript $c$ for simplicity.

According to (\ref{eq:2pt_W}), we have
\begin{align}
    \big\langle
    J(x_1)\,J(x_2)\big\rangle_{W}&=\frac{F(\xi,\eta)}{|\vec{x}_1|^2\,|\vec{x}_2|^2} ~,\qquad
    \big\langle K(x_1)\,K(x_2)\big\rangle_{W}=\frac{G(\xi,\eta)}{|\vec{x}_1|^3\,|\vec{x}_2|^3}
    \label{WJJWKK}
\end{align}
where $F$ and $G$ are functions of the invariants $\xi$ and $\eta$ defined in 
(\ref{eq:CrossRatios}). Here $F$ corresponds to the 0-channel in the R-symmetry expansion from \eqref{eq:Rsymmetry_channel}. To avoid clutter in the notation, we have omitted the subscript 0 since, as discussed above, only the 0-channel can be probed.

Exploiting the SO(3) symmetry, we can rotate the transverse axes so that the points $x_1$ and $x_2$ take the form
\begin{align}
    x_1=(r_1,0,0,t_1)\quad\mbox{and}\quad x_2=(r_2\cos \theta,r_2\sin\theta,0,t_2)
    \label{x1x2}
\end{align}
and the cross-ratios (\ref{eq:CrossRatios}) become
\begin{align}
    \xi=\frac{t_{12}^2+r_1^2+r_2^2}{2\,r_1 r_2}~,\qquad
    \eta=\cos\theta~.
    \label{xieta1}
\end{align}
Then, (\ref{Iis3}) becomes
\begin{equation}
\begin{aligned}
  \mathcal{I}&=-128\pi^2\!\!\int_{-1}^{+1}\!\!d\eta\!\int_0^\infty\!\!dr_1\!\int_{-\infty}^\infty\!\!dt_1\int_0^\infty\!\!dr_2\!\int_{-\infty}^\infty\!\!dt_2\,\frac{F(\xi,\eta)}{(1+r_1^2+t_1^2)^2(1+r_2^2+t_2^2)^2}\\[2mm]
    &\quad+32\pi^2\!\!\int_{-1}^{+1}\!\!d\eta\!\int_0^\infty\!\!dr_1\!\int_{-\infty}^\infty\!\!dt_1\int_0^\infty\!\!dr_2\!\int_{-\infty}^\infty\!\!dt_2\,\frac{G(\xi,\eta)}{r_1 r_2\,(1+r_1^2+t_1^2)
    (1+r_2^2+t_2^2)}
    \label{YWJK2}
\end{aligned}
\end{equation}
with $\xi$ given by (\ref{xieta1}).
Exploiting the SO(1,2) symmetry, we can now bring the points $x_1$ and $x_2$ in (\ref{x1x2}) to a reference configuration
\begin{align}
     x_1=(1,0,0,0)\quad\mbox{and}\quad x_2=(\rho\cos \theta,\rho\sin\theta,0,0)
    \label{x1x2bis}
\end{align}
in which the invariant $\xi$ becomes
\begin{align}
    \xi=\frac{1}{2}\Big(\rho+\frac{1}{\rho}\Big)~,
\end{align}
while $\eta$ remains equal to $\cos\theta$. Thus, we can trade the integrals in (\ref{YWJK2}) for the integrals over the parameters of the SO(1,2) transformation that maps the reference configuration (\ref{x1x2bis}) to the generic one in (\ref{x1x2}). This can be realized by a 
special conformal transformation of parameter $b$ along the Wilson line followed by a dilatation of parameter $\lambda$ and finally by a translation of parameter $a$ along the Wilson line, obtaining
\begin{equation}
    \begin{aligned}
    t_1&=a+\frac{\lambda \,b}{1+b^2}~,\qquad r_1=\frac{\lambda}{1+b^2}~,\\
    t_2&=a+\frac{\lambda \,b \,\rho^2}{1+b^2\rho^2}~,\quad r_2=\frac{\lambda\,\rho}{1+b^2\rho^2}~.
    \label{change}
\end{aligned}
\end{equation}
The Jacobian associated to this change of variables is
\begin{align}
    \Big|\frac{(\partial t_1,\partial r_1,\partial t_2,\partial r_2)}{(\partial\rho,\partial a,\partial b,\partial\lambda)}\Big|=\begin{cases}
       {\displaystyle \frac{(1-\rho^2)\lambda^2}{(1+b^2)^2(1+b^2\rho^2)^2}\quad\mbox{for}~~\rho\leq 1}~,\\[6mm]
       {\displaystyle \frac{(\rho^2-1)\lambda^2}{(1+b^2)^2(1+b^2\rho^2)^2}\quad\mbox{for}~~\rho\geq 1}~.
    \end{cases}
\end{align}

Let us now consider the integral in the first line of (\ref{YWJK2}). After the change of variables (\ref{change}), the integrals over $a$, $b$ and $\lambda$ can be performed analytically and one remains with
\begin{equation}
    \begin{aligned}
    -128\pi^2\!\!\int_{-1}^{+1}\!\!d\eta\Bigg[&-\int_0^1\!\!d\rho\,\frac{\pi ^2 \big(1-\rho ^2+(1+\rho ^2) \log\rho\big)}{2(1-\rho^2)^2}\,F(\xi,\eta)\\
    &+\int_1^\infty\!\!d\rho\,\frac{\pi ^2 \big(1-\rho ^2+(1+\rho ^2) \log\rho\big)}{2(\rho^2-1)^2}\,F(\xi,\eta)
    \Bigg]_{\xi=\frac{1}{2}(\rho+\frac{1}{\rho})}~.
    \label{int1}
\end{aligned}
\end{equation}
Replacing $\rho\to1/\rho$, the second line of (\ref{int1}) becomes equal to the first line and thus the integral involving $F$ becomes
\begin{align}
    128\pi^4\!\!\int_{-1}^{+1}\!\!d\eta\int_0^1\!\!d\rho\,\frac{\big(1-\rho^2+(1+\rho^2) 
    \log\rho\big)}{(1-\rho^2)^2}\,F(\xi,\eta)\Big|_{\xi=\frac{1}{2}(\rho+\frac{1}{\rho})}
\end{align}
In a similar way the integral in the second line of (\ref{YWJK2}) involving the function $G$ can be evaluated to be
\begin{align}
    -64\pi^4\!\!\int_{-1}^{+1}\!\!d\eta\int_0^1\!\!d\rho\,\frac{
    \log\rho}{\rho}\,G(\xi,\eta)\Big|_{\xi=\frac{1}{2}(\rho+\frac{1}{\rho})}~.
\end{align}
In conclusion we have shown that
\begin{align}
    \mathcal{I}&= 
   \partial_m^2 \log \big\langle\!\!\big\langle W_{\mathcal{C}}\big\rangle\!\!\big\rangle_{\mathcal{N}=2^*} \Big|_{m=0}\notag\\[2mm]
   &=128\pi^4\!\!\int_{-1}^{+1}\!\!d\eta\int_0^1\!\!d\rho\,\bigg[\frac{\big(1-\rho^2+(1+\rho^2) 
    \log\rho\big)}{(1-\rho^2)^2}\,F(\xi,\eta)-\frac{
    \log\rho}{2\rho}\,G(\xi,\eta)\bigg]_{\xi=\frac{1}{2}(\rho+\frac{1}{\rho})}~.
    \label{Ifin}
\end{align}
To make further progress we now exploit the supersymmetries preserved by the Wilson line which allow us to establish a relationship between the functions $F$ and $G$ by means of the supersymmetric Ward identities satisfied by the bulk correlators in presence of a Wilson line.

\subsection{Ward identity and final form of the integrated correlators}
In this section we outline the derivation of the Ward identities which allow to relate the functions
$F$ and $G$. A detailed derivation is presented in Appendix~\ref{appendix:ward}.

The first element is the fact that the Wilson line only preserves half of the supersymmetries.
Indeed, $W_l$ is invariant only under transformations whose parameters satisfy the BPS condition
\begin{equation}\label{BPS_WL}
    \bar\xi^i_{\dot\alpha} = \xi^{i\alpha} (\sigma^4)_{\alpha\dot\alpha}~.
\end{equation}
This enforces an effective identification between chiral and anti-chiral spinors, in agreement with the fact that the original rotation symmetry SO(4) of the theory is reduced to SO(3) by the presence of
the line defect. As a consequence of this, it is convenient to repackage the components (\ref{flavorhyper}) of the current multiplet and use the following fields
\begin{align}
    \Phi^{ij}~,~~Y_{i}=X_{i}+\ii\,\overbar{X}_i~,~~ Z_{i}=X_{i}-\ii\,\overbar{X}_i~,~~
    j^m~,~~ S=Q-2\,j^4~,~~T
    \label{newcurrmult}
\end{align}
where
\begin{equation}
    T=P+\overbar{P}~,~~Q=P-\overbar{P}~.
    \label{QTvsPPbar}
\end{equation}
The advantage of using these combinations is that the supersymmetry transformations preserved by the Wilson line take a simpler form (see (\ref{SUSYcurrentBPS}) in Appendix~\ref{appendix:ward}). Moreover, the operator $K$ that appears in the mass deformation coincides with $T$, up to a factor of $-(\ii/2)$.

The second element is that $R$-symmetry, flavor symmetry and conformal symmetry in presence of a line defect severely constrain the form of the correlators. For example, according to (\ref{eq:2pt_W}) for the scalar operators we have
\begin{align}
    \big\langle \Phi^{ij}(x_1)\,\Phi^{k\ell}(x_2)\big\rangle_W&=(\epsilon^{ik}\epsilon^{j\ell}+\epsilon^{i\ell}\epsilon^{jk})\,A(x_1,x_2)~~~\mbox{with}~~~A(x_1,x_2)=\frac{a(\xi,\eta)}{|\vec{x}_1|^2|\vec{x}_2|^2}~,\label{PhiPhi}
    \end{align}
where the index structure in the right hand side is dictated by $R$-symmetry, and
\begin{align}
\big\langle T(x_1)\,T(x_2)\big\rangle_W=H(x_1,x_2)=\frac{h(\xi,\eta)}{|\vec{x}_1|^3|\vec{x}_2|^3}~,~~~
    \big\langle Q(x_1)\,Q(x_2)\big\rangle_W=L(x_1,x_2)=\frac{\ell(\xi,\eta)}{|\vec{x}_1|^3|\vec{x}_2|^3}~.\label{TTQQ}
\end{align}
Also the correlator of two currents in a defect CFT depends on the coordinates and the cross-ratios
in a very specific form. This has been derived in \cite{Lauria:2018klo,Herzog:2020bqw} and is presented in (\ref{jj}).

The third element to consider is that all these structures must be compatible with the supersymmetry transformations of the various fields. For example, starting from the correlators
\begin{align}
\label{WI_constraints_1}
     \big\langle \Phi^{ij}(x_1)\,Y_{k\alpha}(x_2)\big\rangle_W=0~,~~~
     \big\langle S(x_1)\,Y_{i\alpha}^{}(x_2)\big\rangle_W=0~,
\end{align}
which identically vanish because of statistics, and computing their supersymmetry variations one can show that
\begin{align}
\label{eq:S_to_A}
    \big\langle S(x_1)\,S(x_2)\big\rangle_W=-16\,\partial_{4}^{(1)}\partial_4^{(2)}A(x_1,x_2)
\end{align}
where $\partial_\mu^{(a)}$ denotes the derivative with respect to
$x_a^\mu$. In a very similar way, starting from the vanishing correlators 
\begin{equation}
    \big\langle Z_{k\alpha}(x_1),\Phi^{ij}(x_2)\big\rangle_W=0~,~~~
    \big\langle Z_{i\alpha}(x_1)\,S(x_2)\big\rangle_W=0~,
\end{equation}
one finds
\begin{equation}
    \big\langle j^m(x_1)\,S(x_2)\big\rangle_W=8\,\partial_m^{(1)}\partial_4^{(2)}A(x_1,x_2)~.
    \label{eq:jS_to_A}
\end{equation}
Finally, from the supersymmetry variation of
\begin{equation}
    \big\langle Z_{i\alpha}(x_1)\,T(x_2)\big\rangle_W=0~,~~~
    \big\langle Z_{i\alpha}(x_1)\,j^n(x_2)\big\rangle_W=0~,
\end{equation}
one obtains
\begin{equation}
    \big\langle j^m(x_1)\,j^n(x_2)\big\rangle_W=\frac{1}{4}\,
    \big\langle T(x_1)\,T(x_2)\big\rangle_W\,\delta^{mn}-4\,\partial_m^{(1)}\partial_n^{(2)}A(x_1,x_2)~.
    \label{eq:jjH_to_A}
\end{equation}
The details on the derivations of these relations are given in
Appendix~\ref{appendix:ward} where one can find also other identities among correlators which are summarized in Tab.~\ref{table:susyWI}.

By inserting the form of the correlators prescribed by the defect CFT in the relations (\ref{eq:S_to_A}), (\ref{eq:jS_to_A}) and (\ref{eq:jjH_to_A}), one can read how the scalar functions $h(\xi,\eta)$ and $\ell(\xi,\eta)$ introduced in (\ref{TTQQ}) are related to $a(\xi,\eta)$. The result
of this calculation, whose details are reported in Appendix~\ref{appendix:ward}, is
quite simple; in fact
\begin{equation}
    h(\xi,\eta)=-32\,\partial_\eta a(\xi,\eta)~,~~~
    \ell(\xi,\eta)=-32\,\partial_\xi a(\xi,\eta)~.
    \label{hl_to_a}
\end{equation}
Furthermore, we have found that the consistency of the solution of the Ward identities requires that the function $a(\xi,\eta)$ has to satisfy the following
equation
\begin{equation}
    2a+\xi\partial_\xi a+\eta\partial_\eta a=0~.
    \label{homogeneity0}
\end{equation}

We can now collect our findings and determine the relation between the functions $F(\xi,\eta)$ and $G(\xi,\eta)$ that appear in the integrated correlators. Using (\ref{JPhiKP}) and (\ref{PhiPhi}), we have
\begin{align}
    \big\langle J(x_1)\,J(x_2)\big\rangle_W=
    \big\langle \big(\Phi^{11}(x_1)+\Phi^{22}(x_2)\big)\,\big(\Phi^{11}(x_2)+\Phi^{22}(x_2)\big\rangle_W
    =\frac{4\,a(\xi,\eta)}{|\Vec{x_1}|^2|\vec{x}_2|^2}~,
\end{align}
from which we read that
\begin{equation}
    F(\xi,\eta)=4\,a(\xi,\eta)~.
    \label{F_to_a}
\end{equation}
On the other hand, from (\ref{JPhiKP}) and (\ref{TTQQ}) we have
\begin{align}
    \big\langle K(x_1)\,K(x_2)\big\rangle_W=
    \Big(\!\!-\frac{\ii}{2}\Big)^2\,\big\langle T(x_1)\,T(x_2)\big\rangle_W
    =-\frac{1}{4}\frac{h(\xi,\eta)}{|\Vec{x_1}|^3|\vec{x}_2|^3}~,
\end{align}
yielding
\begin{equation}
    G(\xi,\eta)=-\frac{1}{4}\,h(\xi,\eta)~.
    \label{G_to_h}
\end{equation}
Finally, using (\ref{hl_to_a}) we arrive at the relation
\begin{equation}
    G(\xi,\eta)=8\,\partial_\eta a(\xi,\eta)=2\,\partial_\eta F(\xi,\eta)
    \label{G_to_F}
\end{equation}
which implies that (\ref{Ifin}) can be rewritten as follows
\begin{align}
    \mathcal{I}=128\pi^4\!\!\int_{-1}^{+1}\!\!d\eta\int_0^1\!\!d\rho\,\bigg[\frac{\big(1-\rho^2+(1+\rho^2) 
    \log\rho\big)}{(1-\rho^2)^2}\,F(\xi,\eta)-\frac{
    \log\rho}{\rho}\,\partial_\eta F(\xi,\eta)\bigg]_{\xi=\frac{1}{2}(\rho+\frac{1}{\rho})}~.
    \label{Ifin1}
\end{align}
Since the second term in the square brackets is a total derivative, it contributes only to boundary terms and thus, according to the observation in footnote~\ref{footnote_boundary}, it can be discarded.

\section{The matrix model calculation}
\label{sec:matrix}
We now present the calculation of the integrated correlator $\mathcal{I}$ using the matrix model approach. Even if this calculation has already appeared in the literature \cite{Pufu:2023vwo} (see also \cite{Russo:2013kea}), we propose an alternative derivation based on
the use of recursion relations and Bessel kernels \cite{Billo:2022fnb} which is very direct and may be useful also for generalizations to massive deformations of $\mathcal{N}=2$ theories.

As is well known, using localization techniques the $\mathcal{N}=2^*$ SYM theory on a 4-sphere can be described by a matrix model \cite{Pestun:2007rz} whose partition function is
\begin{align}
   \mathcal{Z}(m)=\int \!da~\rme^{-\frac{8\pi^2N}{\lambda}\,\tr\,a^2}\,\Big|Z_{\mathrm{1-loop}}(a,m)
\,Z_{\mathrm{inst}}(a,\lambda,m)\Big|^2~.
\label{Z2*}
\end{align}
Here $a$ is a Hermitian matrix in the fundamental representation of SU($N$) which we write as
\begin{equation}
a=\sum_{b=1}^{N^2-1} a^b\,T_b
\end{equation}
where the generators are normalized in such a way that $
\tr\,T_b\,T_c =\frac{1}{2}\,\delta_{bc}$. 
In the following, use the so called ``full-Lie algebra approach'', introduced in \cite{Billo:2017glv}, namely we integrate over all matrix elements of $a$ with the normalized measure
\begin{equation}
da=\prod_{b=1}^{N^2-1}\frac{da^b}{\sqrt{2\pi}}~.
\end{equation}
The two factors $Z_{\mathrm{1-loop}}(a,m)$ and $Z_{\mathrm{inst}}(a,\lambda,m)$ are the contributions of the 1-loop and instanton fluctuations around the localization locus. In the large-$N$ limit we can neglect instanton effects and set $Z_{\mathrm{inst}}(a,\lambda,m)=1$.
The explicit expression of the 1-loop term, which does not depend on the 't Hooft coupling $\lambda$, is given in \cite{Pestun:2007rz}
and can be regarded as an interaction action in the matrix model, namely
\begin{align}
\Big|Z_{\mathrm{1-loop}}(a,m)\Big|^2=\rme^{-S_{\mathrm{int}}(a,m)}~.
\end{align}
In the limit of small $m$, we have
\begin{align}
S_{\mathrm{int}}(a,m)=-\frac{m^2}{2} \bigg[\sum_{n=1}^\infty\sum_{\ell=0}^{2n}(-1)^{n+\ell}\frac{(2n+1)!}{\ell! (2n-\ell)!}\,\zeta_{2n+1}\tr\,a^{2n-\ell}\,\tr\,a^{\ell}\bigg]+
O(m^4)
\end{align}
where $\zeta_k$ is the Riemann-$\zeta$ value $\zeta(k)$. From the partition function, we obtain
\begin{equation}
    \partial_m^2\log \mathcal{Z}(m)\Big|_{m=0}=\frac{1}{\mathcal{Z}(0)}\,\int \!da~\rme^{-\frac{8\pi^2N}{\lambda}\tr\,a^2}
    \bigg[\sum_{n=1}^\infty\sum_{\ell=0}^{2n}(-1)^{n+\ell}\frac{(2n+1)!}{\ell! (2n-\ell)!}\,\zeta_{2n+1}\tr\,a^{2n-\ell}\,\tr\,a^{\ell}\bigg]~.
\end{equation}
Rescaling the matrix $a$ according to
\begin{align}
    a\,\to\,\sqrt{\frac{\lambda}{8\pi^2N}}\,a~,
    \label{rescaling}
\end{align}
we have
\begin{align}
  \partial_m^2\log \mathcal{Z}(m)\Big|_{m=0}=  
  \int \!da~\rme^{-\tr\,a^2}\, \mathbf{M}(a,\lambda)
\,\equiv\,\big\langle \mathbf{M}(a,\lambda) \big\rangle_0
\label{d2mZ}
\end{align}
where
\begin{align}
\mathbf{M}(a,\lambda)=-\sum_{n=1}^\infty\sum_{\ell=0}^{2n}(-1)^{n+\ell}\frac{(2n+1)!}{\ell! (2n-\ell)!}\,\zeta_{2n+1}\Big(\frac{\lambda}{8\pi^2N}\Big)^n\tr\,a^{2n-\ell}\,\tr\,a^{\ell}~,
\label{Xla}
\end{align}
and the notation $\langle~\rangle_0$ stands for the vacuum
expectation value in the free Gaussian matrix model.

It is now useful to change basis and rewrite $\mathbf{M}(a,\lambda)$
in terms of the normal-ordered operators $\mathcal{P}_n(a)$ introduced in \cite{Beccaria:2021hvt,Billo:2022fnb}\,%
\footnote{In \cite{Beccaria:2021hvt} these operators were denoted by $\omega_n(a)$.} which satisfy the following properties
\begin{equation}
    \label{2ptP}
        \big\langle \mathcal{P}_n(a)\big\rangle_0 = 0~,\qquad\big\langle \mathcal{P}_m(a)\, \mathcal{P}_n(a)\big\rangle_0 = \delta_{mn}~,
\end{equation}
and are related to $\tr a^n$ as follows:
\begin{equation}
   \label{atoP}
    \tr a^n = 
    \Big(\frac{N}{2}\Big)^{n/2} 
    \sum_{k=0}^{\left[\frac{n-1}{2}\right]}
    \sqrt{n-2k} \binom{n}{k} \,\mathcal{P}_{n-2k}(a)+
    \big\langle \tr a^n\big\rangle_0~.
\end{equation}
Inserting this relation in (\ref{Xla}), with some simple algebra one can prove that
\begin{equation}
    \label{X012}
        \mathbf{M}(a,\lambda) = \mathbf{M}^{(0)}(\lambda) + \mathbf{M}^{(1)}(a,\lambda) + \mathbf{M}^{(2)}(a,\lambda)
\end{equation}
where the three terms containing, respectively, zero, one and two normal-ordered operators\,%
\footnote{Notice that in the planar 't Hooft limit $\mathbf{M}^{(0)}$ is of order $N^2$, $\mathbf{M}^{(1)}$ is of order $N$ and $\mathbf{M}^{(2)}$ is of order $N^0$.}, can be expressed as follows
\begin{align}
    \label{X012are}
        \mathbf{M}^{(0)}(\lambda) & =  N^2 \,\mathsf{M}_{0,0}(\lambda)~,\notag\\[1mm]
        \mathbf{M}^{(1)}(a,\lambda) &=2N\, \sum_{k=1}^\infty 
        \mathsf{M}_{0,2k}(\lambda)\, \mathcal{P}_{2k}(a)~,\notag\\
        \mathbf{M}^{(2)}(a,\lambda)&=  \sum_{k,\ell=1}^\infty 
        \Big[\mathsf{M}_{2k,2\ell}(\lambda)\, \mathcal{P}_{2k}(a) \,\mathcal{P}_{2\ell}(a)+
        \mathsf{M}_{2k+1,2\ell+1}(\lambda)\, \mathcal{P}_{2k+1}(a) \,\mathcal{P}_{2\ell+1}(a)\Big] ~.
\end{align}
Here the $\lambda$-dependent coefficients are given by the following convolutions of Bessel functions of the first kind $J_r$:
\begin{align}
\mathsf{M}_{0,0}(\lambda)&=\int_0^\infty \!\frac{dx}{x}~
        \chi\Big(\frac{2\pi x}{\sqrt{\lambda}}\Big)
       \bigg[1-
        \Big(\frac{2 J_1(x)}{x}\Big)^2
        \bigg]~,\notag\\[2mm]
\mathsf{M}_{0,r}(\lambda) & = (-1)^{\frac{r}{2}+1} \,\sqrt{r}\int_0^\infty \!\frac{dx}{x}~
        \chi\Big(\frac{2\pi x}{\sqrt{\lambda}}\Big) \, 
        \frac{2J_1(x)}{x}\,J_{r}(x)~~~(r\geq 1)~,\notag\\[2mm]
\mathsf{M}_{r,s}(\lambda)&=(-1)^{\frac{r+s+2rs}{2}+1}\, \sqrt{r\,s} \int_0^\infty \!\frac{dx}{x}~ \chi\Big(\frac{2\pi x}{\sqrt{\lambda}}\Big) \, J_{r}(x)\,J_{s}(x)~~~(r,s\geq 1)~,
\label{Mrs}
\end{align}
where the kernel function is
\begin{equation}
    \chi(x)=\frac{\big(x/2\big)^2}{\sinh^2\!\big(x/2\big)}~.
    \label{chifv}
\end{equation}
We remark that even if we started from a perturbative expansion in powers of $\lambda$ (see (\ref{Xla})), using this formalism we have found expressions that can be continued beyond perturbation theory and hold for any value of $\lambda$. Indeed, the integrals in (\ref{Mrs}) can be easily studied at strong coupling where the asymptotic expansion for $\lambda\to\infty$ can be obtained with a straightforward application of the Mellin-Barnes method. It is interesting to observe that the coefficients $\mathsf{M}_{r,s}(\lambda)$ in (\ref{Mrs}) have the same structure of the elements $\mathsf{X}_{r,s}(\lambda)$ appearing in the study of the 2- and 3-point functions of scalar operators in certain $\cN=2$ superconformal gauge theories \cite{Beccaria:2021hvt,Billo:2021rdb,Billo:2022gmq,Billo:2022xas,Billo:2022fnb,Billo:2022lrv}. The only difference is in the kernel function $\chi(x)$, which in the present case is given in (\ref{chifv}), whereas there is equal to $2/\sinh^2\!\big(x/2\big)$. Similar convolutions of Bessel functions with different kernels appear also in the study of other observables
both in $\cN=4$ and in $\cN=2$ SYM theories (see for example \cite{Beccaria:2022ypy} and references therein).

Using (\ref{X012are}) in (\ref{d2mZ}), it is now immediate to see that
\begin{equation}
    \label{Xis}
        \partial_m^2\log \mathcal{Z}(m)\Big|_{m=0} = N^2  
        \int_0^\infty\! \frac{dx}{x}~
        \chi\Big(\frac{2\pi x}{\sqrt{\lambda}}\Big)
       \bigg[1-
        \Big(\frac{2 J_1(x)}{x}\Big)^2
        \bigg]+ \dots
\end{equation}
where the ellipses stand for terms subleading in the large-$N$ limit, in agreement with the results of \cite{Russo:2013kea}. We note in passing that by applying the hyperbolic Laplacian operator $\Delta_\tau=\tau_2^2\partial_\tau\partial_{\overbar{\tau}}$, where
\begin{equation}
    \tau=\tau_1+\ii\,\tau_2=\frac{\theta}{2\pi}+\ii\,\frac{4\pi N}{\lambda}
\end{equation}
is the standard complexified Yang-Mills coupling, we readily obtain
\begin{align}
    \Delta_\tau \,\partial_m^2\log \mathcal{Z}(m)\Big|_{m=0} &=\frac{N^2}{2}\big[\mathsf{M}_{2,2}(\lambda)-2\mathsf{M}_{1,1}(\lambda)\big]+\dots\notag\\[1mm]
    &=N^2\int_0^\infty\! \frac{dx}{x}~
        \chi\Big(\frac{2\pi x}{\sqrt{\lambda}}\Big)
       \big[J_1(x)^2-J_2(x)^2\big]+\dots~,
\end{align}
which agrees with the expression derived in \cite{Binder:2019jwn,Dorigoni:2021guq} in the study of the integrated correlator of four scalar operators in $\mathcal{N}=4$ SYM.

Let us now introduce in the matrix model the circular Wilson loop that, after the rescaling (\ref{rescaling}), is represented by the following operator \cite{Pestun:2007rz}
\begin{equation}
\mathbf{W}(a,\lambda)=\frac{1}{N}\,\tr\,\exp\Big(\sqrt{\frac{\lambda}{2N}}\,a\Big)
~.
\label{WL}
\end{equation}
In the large $N$-limit we can rewrite it in terms of the normal-ordered operators as follows
\begin{align}
    \label{WPn0}
        \mathbf{W}(a,\lambda)= \big\langle
        \mathbf{W}(a,\lambda)\big\rangle_0 + \sum_{n=2}^\infty w_n(\lambda)\, \mathcal{P}_n(a)
\end{align}
where $\big\langle\mathbf{W}(a,\lambda)\big\rangle_0$ is the vacuum expectation value in the free matrix model given by the well-known result in terms of the modified Bessel function \cite{Erickson:2000af}
\begin{align}
    \big\langle
        \mathbf{W}(a,\lambda)\big\rangle_0=\frac{2\,I_1(\sqrt{\lambda})}{\sqrt{\lambda}}~,
        \label{W0}
\end{align}
and the $\lambda$-dependent coefficients $w_n$ are
\begin{align}
    \label{checkWP}
        w_n(\lambda)=\big\langle \mathbf{W}(a,\lambda)\, \cP_n(a)\big\rangle_0 = N\,\sqrt{n}\,I_n(\sqrt{\lambda})~,
\end{align}
as one can check using Eq.~(4.30) of \cite{Billo:2018oog}.

Let us now consider the vacuum expectation value of the Wilson loop operator (\ref{WPn0}) in the massive $\mathcal{N}=2^*$ theory. This 
vacuum expectation value, which is the matrix-model counterpart of the gauge theory expression
(\ref{vevW}), reads
\begin{align}
    \mathcal{W}(m)=\frac{1}{\mathcal{Z}(m)}\,\int\!
    da~\mathbf{W}(\lambda,a)\,\,\rme^{-\tr a^2}\,\rme^{\frac{m^2}{2}\,\mathbf{M}(\lambda,a)+O(m^4)}~,
\end{align}
so that
\begin{align}
    \mathcal{I}&=\partial_m^2\log \mathcal{W}(m)\Big|_{m=0}=
    \frac{\big\langle \mathbf{W}(a,\lambda)\,\mathbf{M}(a,\lambda)\big\rangle_0-\big\langle \mathbf{W}(a,\lambda)\big\rangle_0\,\big\langle\mathbf{M}(a,\lambda)\big\rangle_0}{\big\langle \mathbf{W}(a,\lambda)\big\rangle_0}~.
\end{align}
In the large-$N$ limit, we can use (\ref{X012}), (\ref{X012are}), (\ref{Mrs}), (\ref{WPn0}) and (\ref{W0}) to obtain
\begin{align}
    \mathcal{I}&=\frac{\sqrt{\lambda}}{2NI_1(\sqrt{\lambda})}\,\sum_{n=2}^\infty\sqrt{n}\,I_n(\sqrt{\lambda})\,\big\langle
    \mathcal{P}_n(a)\,\mathbf{M}^{(1)}(a,\lambda)\big\rangle_0
    +\dots\notag\\[1mm]
    &=\frac{\sqrt{\lambda}}{I_1(\sqrt{\lambda})}\,\sum_{k=1}^\infty\sqrt{2k}\,I_{2k}(\sqrt{\lambda})\,\mathsf{M}_{0,2k}(\lambda)
    +\dots\notag\\[1mm]
&=-\frac{\sqrt{\lambda}}{I_1(\sqrt{\lambda})}\,\int_0^\infty\!\frac{dx}{x}~\chi\Big(\frac{2\pi x}{\sqrt{\lambda}}\Big)\,\frac{2J_1(x)}{x}\sum_{k=1}^\infty(-1)^k \,2k\, I_{2k}(\sqrt{\lambda})\,J_{2k}(x)+\dots~.
    \label{Im}
\end{align}
Using the identity
\begin{equation}
\label{ida}
    \sum_{k=1}^\infty (-1)^k\, 2k\, I_{2k}(\sqrt{\lambda})\, J_{2k}(x) = - \frac{\sqrt{\lambda}\, x}{2\,(x^2+ \lambda)}\, \Big[
    \sqrt{\lambda} \,I_0(\sqrt{\lambda})\, J_1(x)-x\, I_1(\sqrt{\lambda})\, J_0(x) \Big]~,
\end{equation}
which simply follows from the recursion properties of the Bessel functions,
we finally obtain
\begin{equation}
\label{Yisa}
\mathcal{I}= \frac{\lambda}{I_1(\sqrt{\lambda)}}\,
\int_0^\infty\! \frac{dx}{x}~\chi\Big(\frac{2\pi x}{\sqrt{\lambda}}\Big)\,\frac{\sqrt{\lambda}\,I_0(\sqrt{\lambda})\,J_1(x)^2-x\,I_1(\sqrt{\lambda})\,J_0(x)\,J_1(x)}{x^2+\lambda}+\dots~.
\end{equation}
This agrees with the results of \cite{Russo:2013kea} (see
also \cite{Pufu:2023vwo})\,%
\footnote{Note that in \cite{Russo:2013kea} the result is not normalized with respect to the vacuum expectation value (\ref{W0}), and that we differ by an overall normalization factor of 1/2 with respect to \cite{Pufu:2023vwo}.}. Expanding (\ref{Yisa}) for small values of $\lambda$, it is straightforward to obtain the perturbative expansion of $\mathcal{I}$, whose first few terms are
\begin{equation}
    \mathcal{I}~\underset{\lambda \rightarrow 0}{\sim}~\frac{3\,\zeta_3}{32\pi^2}\,\lambda^2-\Big(\frac{\zeta_3}{256\pi^2}+\frac{25\,\zeta_5}{256\pi^4}\Big)\lambda^3+\Big(\frac{\zeta_3}{4096 \pi^2}+\frac{15\, \zeta_5}{4096 \pi^4}+\frac{735\,\zeta_7}{8192 \pi^6}\Big)\lambda^4+O(\lambda^5)+\dots~.
    \label{Ipert}
\end{equation}
On the other hand, using the asymptotic behavior of the Bessel functions for large values of $\lambda$, we can easily compute from (\ref{Yisa}) the strong-coupling expansion of $\mathcal{I}$, whose first terms are
\begin{equation}
    \mathcal{I}~\underset{\lambda \rightarrow \infty}{\sim}~
    \frac{\sqrt{\lambda}}{2}+\Big(\frac{1}{4}-\frac{\pi^2}{6}\Big)+
    O(\lambda^{-1/2})+\dots~.
    \label{Istrong}
\end{equation}

\section{Conclusions and open problems}
\label{sec:conclusions}
Our main result is the integral relation (\ref{Ifin1}) 
between the second mass derivative of the Wilson loop expectation value and the function $F$ that appears in the correlator of two specific scalar operators belonging to the stress-tensor multiplet
in presence of a line defect.
The derivation of this result heavily relies on the analysis of the constraints imposed on correlators by the symmetries of the $\cN=4$ SYM theory that survive
when the mass deformation is introduced and a $\frac{1}{2}$-BPS Wilson line is added. Quite remarkably, the superconformal Ward identities that follow from these constraints admit a solution in terms of a single function of the two invariant cross-ratios that enter in the correlation functions of two bulk operators in presence of a line defect. Therefore, as shown at the end of Section~\ref{sec:integrated} (see in particular 
(\ref{hl_to_a})--(\ref{G_to_F})), all contributions to the mass derivative of the expectation value of the Wilson loop can be expressed as integrals of this function with a specific integration measure. Quite unexpectedly, however, in solving the superconformal Ward identities among 2-point functions of bulk operators, we found that this function  has to satisfy a homogeneity condition (see (\ref{homogeneity0}) or (\ref{omogeneity})). It would be very interesting to explore the implications of this condition, and more generally to carry out a detailed analysis of the perturbative expansion of the integrated correlators with a Wilson loop and compare it with the matrix model predictions, similarly to what has been done in \cite{Wen:2022oky} for the integrated 4-point functions. Another open problem is the study of the properties of the integrated correlators at strong coupling from the point of view of field theory, also in relation to the explicit results of the matrix model derived in \cite{Pufu:2023vwo} and briefly reviewed in Section~\ref{sec:matrix}.

\vskip 1cm
\noindent {\large {\bf Acknowledgments}}
\vskip 0.2cm
We would like to thank Lorenzo Bianchi, Alessandro Georgoudis, Luca Griguolo, Gregory Korchemsky, Marco Meineri, Alessandro Pini, Oliver Schlotterer and Kostantin Zarembo for many useful discussions. We also thank Kostantin Zarembo, Luca Griguolo and the authors of \cite{Pufu:2023vwo} for reading and commenting a preliminary version of this paper.
This research is partially supported by the MUR PRIN contract 2020KR4KN2 ``String Theory as a bridge between Gauge Theories and Quantum Gravity'' and by
the INFN project ST\&FI
``String Theory \& Fundamental Interactions''. \\
FG acknowledges support from the Simons Foundation grant 994306 (Simons Collaboration on Confinement and QCD Strings) and from NCCR SwissMAP.

\vskip 1cm

\appendix

\section{Conventions}
\label{appendix:spinor}
\subsection*{Indices}
Our conventions for indices are the following:
\begin{itemize}
\item[-]
Vector indices in $\mathbb{R}^4$: $\mu,\nu= 1,\dots,4$. 

\item[-] After the line insertion along $x^4$, they split as $\mu=(m,4)$ where $m=1,2,3$ labels the directions orthogonal to the defect.

\item[-]
Chiral and anti-chiral spinor indices in $\mathbb{R}^4$: $\alpha,\beta =1,2$ and $\dot \alpha,\dot \beta =1,2$.

\item[-] After the line insertion, the unique SU$(2)$ spinor index is labeled by $\alpha,\beta =1,2$.

\item[-]
SO$(6)_R$ $R$-symmetry indices: $I,J= 1,\dots 6$.

\item[-]
SU$(2)_R$ $R$-symmetry indices: $i,j= 1,2$.

\item[-]
SU$(2)_F$ flavor symmetry indices: $a,b= 1,2$.

\end{itemize}

\subsection*{Spinors}
We denote by $\psi$ a chiral spinor of components $\psi_\alpha$ with $\alpha=1,2$, and by $\overbar{\psi}$ an anti-chiral spinor with components $\overbar{\psi}^{\dot\alpha}$ with $\dot\alpha=1, 2$. Spinor indices are raised and lowered with the following rules
\begin{equation}
    \psi^\alpha=\psi_\beta\,\epsilon^{\beta\alpha}~,\quad
    \psi_\alpha=\psi^{\beta}\,\epsilon_{\beta\alpha}~,\quad
    \overbar{\psi}^{\dot\alpha}=\overbar{\psi}_{\dot\beta}\,\epsilon^{\dot\beta\dot\alpha}~,\quad
    \overbar{\psi}_{\dot\alpha}=\overbar{\psi}^{\dot\beta}\,\epsilon_{\dot\beta\dot\alpha}
\end{equation}
where
\begin{equation}
    \epsilon^{12}=\epsilon^{\dot 1 \dot 2}=\epsilon_{21}=\epsilon_{\dot 2\dot 1}=1~,
\end{equation}
and they are contracted as follows
\begin{equation}
    \psi \,\chi\,\equiv\,\psi^\alpha\,\chi_\alpha~,\quad
    \overbar{\psi} \, \overbar{\chi}\,\equiv\, 
    \overbar{\psi}_{\dot\alpha}\overbar{\chi}^{\dot\alpha}~.
\end{equation}
Unless necessary to avoid ambiguities, most of the times we will not write explicit indices in the spinor contractions.

We realize the Clifford algebra of $\mathbb{R}^4$ with the matrices $(\sigma^\mu)_{\alpha\dot\beta}$ and $(\overbar{\sigma}^\mu)^{\dot\alpha\beta}$ where
\begin{equation}
    \sigma^\mu=(\Vec{\sigma},-\ii \mathbb{1})~,\quad
    \overbar{\sigma}^\mu=(-\vec{\sigma},-\ii \mathbb{1})
\end{equation}
with $\vec{\sigma}$ being the Pauli matrices. The $\mathfrak{so}(4)$ generators $J^{\mu\nu}=-J^{\nu\mu}$ act on
chiral and anti-chiral spinors as follows
\begin{equation}
    \big(J^{\mu\nu}\big)_\alpha^{~\beta}\, \psi_\beta\qquad \mbox{and}\qquad
    \big(J^{\mu\nu}\big)^{\dot\alpha}_{~\dot\beta}\, \overbar{\psi}^{\dot\beta}
\end{equation}
where
\begin{subequations}
    \begin{align}
    \big(J^{\mu\nu}\big)_\alpha^{~\beta} 
    &= \frac{\ii}{2} \big(\sigma^{\mu\nu}\big)_\alpha^{~\beta} 
    =  \frac{\ii}{8} \big(\sigma^\mu\bar\sigma^\nu - \sigma^\nu\bar\sigma^\mu\big)_\alpha^{~\beta}~,\\
    \big(J^{\mu\nu}\big)^{\dot\alpha}_{~\dot\beta} 
    &= \frac{\ii}{2} \big(\bar\sigma^{\mu\nu}\big)^{\dot\alpha}_{~\dot\beta} 
    =  \frac{\ii}{8} \big(\bar\sigma^\mu\sigma^\nu - \bar\sigma^\nu\sigma^\mu\big)^{\dot\alpha}_{~\dot\beta}~. 
    \end{align}
    \label{Jmunu}%
\end{subequations}
Writing $\mathfrak{so}(4)$ as $\mathfrak{su}(2)_+\oplus\mathfrak{su}(2)_-$, the generators $J^m_{\pm}$
($m=1,2,3$) of $\mathfrak{su}(2)_\pm$ are given by
\begin{align}
    \label{genpm}
        J^m_\pm = \frac 12 \epsilon^{mnp} J^{np} \pm J^{m4}~,
\end{align}
so that, using (\ref{Jmunu}), one finds
\begin{subequations}
\begin{align}
    \big(J^m_+\big)_\alpha^{~\beta} &= \frac{1}{2}(\sigma^m\big)_\alpha^{~\beta}~,~~~
    \big(J^m_-\big)_\alpha^{~\beta} = 0~,\\
    \big(J^m_-\big)^{\dot\alpha}_{~\dot\beta} &= \frac{1}{2}(\sigma^m\big)^{\dot\alpha}_{~\dot\beta}~,~~~
    \big(J^m_+\big)^{\dot\alpha}_{~\dot\beta} = 0~.
\end{align}
\end{subequations}
This shows that a chiral spinor $\psi$ transforms in the $(\mathbf{2},\mathbf{1})$ representation of
$\mathfrak{su}(2)_+\oplus\mathfrak{su}(2)_-$, while an anti-chiral spinor $\overbar{\psi}$ transforms in the $(\mathbf{1},\mathbf{2})$.

\subsection*{Defect spinors}
\label{subappendix:defectspinors}
In presence of a Wilson line along the direction $x^4$, the rotational symmetry $\mathrm{SO}(4)$ is broken to the SO$(3)$ symmetry that rotates the coordinates $x^m$ transverse to the defect. 
At the level of the covering spin group, this breaking corresponds to
\begin{align}
    \label{breakingspinor}
        \mathrm{SU}(2)_+ \times \mathrm{SU}(2)_- \to\,
        \diag\Big(\mathrm{SU}(2)_+ \times \mathrm{SU}(2)_-\!\Big)~.
\end{align}  
One can easily see this also by considering the generators (\ref{genpm}): indeed, in presence of the defect, the rotations $J^{i4}$ are broken, and thus
\begin{align}
    \label{Jpmto}
        J^m_\pm \to J^m =  \frac 12 \epsilon^{mnp} J^{np}~.
\end{align}
Under the breaking pattern (\ref{breakingspinor}), the vector representation 
decomposes as $\mathbf{4}\to \mathbf{3} \oplus \mathbf{1}$,
corresponding to the splitting of a vector $v^\mu$ into $(\vec v,v^4)$.	Similarly, all tensor fields can be decomposed. As far as spinors are concerned, both chiral and anti-chiral spinors collapse to the same spinor representation:
\begin{equation}
    (\mathbf{2},\mathbf{1}) \to \mathbf{2}~,\quad
    (\mathbf{1},\mathbf{2}) \to \mathbf{2}~.
\end{equation}
This means that under rotations of the three-dimensional space transverse to the defect, a chiral spinor with a lower index $\alpha$ behaves in the same way as an anti-chiral spinor with an upper
index $\dot\alpha$. In other words, in presence of the defect we can identify the two indices and simply denote them by $\alpha$.

In our set-up, however, we deal with spinors that carry also an index $i$ of the SU$(2)_R$ symmetry group and satisfy the pseudo-reality conditions
\begin{equation}
    (\psi^i_\alpha)^*\,\equiv\,\psi_i^\alpha=\psi^j_\beta\,\epsilon_{ji}\,\epsilon^{\beta\alpha}\qquad\mbox{and}\qquad
    (\overbar{\psi}^{i\dot\alpha})^*\,\equiv\,\overbar{\psi}_{i\dot\alpha}=
    \overbar{\psi}^{j\dot\beta}\,\epsilon_{ji}\,\epsilon_{\beta\alpha}~.
\end{equation}
Explicitly, these conditions read
\begin{equation}
    \label{Majcondpsi}
    (\psi^1_1)^* = -\psi^2_2~,~~~ (\psi^1_2)^* = \psi^2_1~,
\end{equation}	   
(the signs remain the same if the chiral index is raised), and
\begin{equation}
\label{Majcondpsibar}
    (\overbar{\psi}^{1\dot 1})^* =  \overbar{\psi}^{2\dot 2}~,~~~
    (\overbar{\psi}^{1\dot 2})^* =  -\overbar{\psi}^{2\dot 1}
\end{equation}	
(again, the signs are the same if the anti-chiral index is lowered).
Comparing (\ref{Majcondpsi}) and (\ref{Majcondpsibar}) we see that there is an overall sign difference in the behavior with respect to complex conjugation between the chiral and anti-chiral spinors. Therefore, in the reduction to the defect symmetric notation, we have to set
\begin{equation}
  \label{redcM}
  	\overbar{\psi}^{i\dot\alpha} \to \ii\, \overbar{\psi}^i_\alpha
\end{equation}	 		
with $\bar\psi^i_\alpha$ obeying the pseudo-reality condition (\ref{Majcondpsi}).

We can summarize this discussion by providing a set of effective rules to implement on spinors the defect symmetry breaking pattern:

\begin{equation}
    \begin{aligned}
    \label{rulesds}
         \overbar{\psi}^{i\dot\alpha}&\to \ii \,\overbar{\psi}^i_\alpha~,\hspace{3cm}
        \overbar{\psi}^i_{\dot\alpha}\to -\ii\, \overbar{\psi}^{i\alpha}~,\\
        (\sigma^m)_{\alpha\dot\beta} & \to (\sigma^m)_\alpha^{~\beta}~,\hspace{1.8cm}
        (\overbar{\sigma}^m)^{\dot\alpha\beta} \to -(\sigma^m)_\alpha^{~\beta}~~~(m=1,2,3)~,
        \\
        (\sigma^4)_{\alpha\dot\beta} &\to -\ii\, \delta_\alpha^{~\beta}~,\hspace{2.1cm}
        (\overbar{\sigma}^4)_{\dot\alpha\beta} \to -\ii\, \delta_\alpha^{~\beta}~,\\
       (\sigma^{mn})_{\alpha}^{~\beta} &\to -\ii\,\epsilon^{mnp}\, (\sigma^{p})_{\alpha}^{~\beta}~,\hspace{.5cm}
        (\overbar{\sigma}^{mn})^{\dot\alpha}_{~\dot\beta}  \to -\ii\,\epsilon^{mnp}\, (\sigma^{p})_{\alpha}^{~\beta}~~~(m,n,p=1,2,3)~,\\
        (\sigma^{m4})_{\alpha}^{~\beta} &\to -\ii\, (\sigma^{m})_{\alpha}^{~\beta}~,\hspace{1.3cm}
        (\overbar{\sigma}^{m4})^{\dot\alpha}_{~\dot\beta} \to \ii\, (\sigma^{m})_{\alpha}^{~\beta}~~~(m=1,2,3)~,\\
        \epsilon^{\dot\alpha\dot\beta} &\to - \epsilon_{\alpha\beta}~,\hspace{2.8cm}
         \epsilon_{\dot\alpha\dot\beta} \to - \epsilon^{\alpha\beta}~.
\end{aligned}
\end{equation}

\section{Supersymmetry transformations and multiplets}
\label{appendix:susy}
In this appendix we collect the supersymmetry transformations of the various $\cN=2$ multiplets that were considered in the main text. We denote the supersymmetry parameters by $\xi^{i\alpha}$ and $\overbar{\xi}^i_{\,\dot\alpha}$ with $i=1,2$. The chiral and anti-chiral indices $\alpha$ and $\dot\alpha$ are lowered and raised according to the rules described in Appendix~\ref{appendix:spinor}.
We will only write the supersymmetry transformations for the free theory since this is enough for our purposes. In particular this means that all derivatives are normal non-covariant derivatives. 

\subsection*{The free $\cN=2$ vector multiplet}

This multiplet consists of a vector $A_\mu$, two scalars $\varphi$ and $\overbar{\varphi}$, two chiral fermions $\lambda_{i\alpha}$ and
two anti-chiral fermions $\overbar{\lambda}_i^{\,\dot\alpha}$.
To close the supersymmetry algebra off-shell one has to introduce three auxiliary fields $D_{ij}=D_{ji}$. The $\cN=2$ supersymmetry
transformations are (suppressing spinor indices)
\begin{subequations}
\begin{align}
&\delta A_\mu =\ii\,\xi^{i}\sigma_\mu \overbar{\lambda}_{i}-\ii\,\bar{\xi}^{i}\overbar{\sigma}_\mu \lambda_{i}~,\\[1mm]
&\delta\phi=\sqrt{2} \,\xi^{i}\lambda_{i}~,\quad
\delta\overbar{\phi}=\sqrt{2}\,\bar{\xi}^i\overbar{\lambda}_i~,\\[1mm]
&\delta\lambda_{i}=-\frac{1}{2}\,\xi^{j}\epsilon_{ji}\,\sigma^{\mu\nu}
F_{\mu\nu}-\ii\,\sqrt{2}\,
\bar{\xi}^{j}\epsilon_{ji}\,\overbar{\sigma}^\mu \partial_\mu\phi+\xi^{j}\,D_{ji}~,
\\[1mm]
&\delta\overbar{\lambda}_{i}=-\frac{1}{2}\,
\bar{\xi}^{j}\epsilon_{ji}\,\overbar\sigma^{\mu\nu} F_{\mu\nu}+\ii\,\sqrt{2}\,\xi^{j}\epsilon_{ji}\,\sigma^\mu\partial_\mu\overbar{\phi}+
\bar{\xi}^{j}\,D_ {ji}~,\\[1mm]
&\delta D_{ij}=\ii\,\bar{\xi}^k\epsilon_{ki}\,
\overbar{\sigma}^\mu
\partial_\mu\lambda_{j}-\ii\,\xi^k\epsilon_{ki}\,\sigma^\mu\partial_\mu
\overbar{\lambda}_j+(i\leftrightarrow j)
\end{align}
\label{susyvector}%
\end{subequations}
where $F_{\mu\nu}$ is the field strength of $A_\mu$. These transformations close the following $\cN=2$ algebra
\begin{align}
    \big[\delta_1\,,\,\delta_2\big] =
    -2\,\ii\,(\xi_1^i\,\sigma^\mu \bar{\xi}_2^j-\xi^i_2\,\sigma^\mu\bar{\xi}_1^j)\,\epsilon_{ji}\,\partial_\mu~.
    \label{SUSYalgebra}
\end{align}
One can easily verify that the connection appearing in the Wilson line $W_l$ considered in the main text, namely
\begin{equation}
   \cL= \ii\,A_4+\frac{\varphi+\overbar{\varphi}}{\sqrt{2}}~,
   \label{Wilsonconn}
\end{equation}
is left invariant by supersymmetry transformations in which
\begin{align}
    \bar\xi^i_{\,\dot\alpha}=\xi^{i\alpha}(\sigma^4)_{\alpha\dot\alpha}~.
    \label{BPS}
\end{align}
Thus, the presence of the Wilson line $W_l$ reduces by half the number of supersymmetries and enforces an identification between
chiral and anti-chiral spinors, as explained in Appendix~\ref{subappendix:defectspinors}. The supersymmetries that satisfy the condition (\ref{BPS}) close the following algebra
\begin{align}
    \big[\delta_1\,,\,\delta_2\big] =
    -4\,\ii\,\xi_1^{i}\,{\xi}_2^j\,\epsilon_{ji}\,\partial_4~,
    \label{SUSYalgebraBPS}
\end{align}
in agreement with the fact that only the translations in the direction of the line defect ($x^4$ in our case) are a symmetry of the configuration.

Applying to (\ref{BPS}) the effective rules (\ref{rulesds}), we easily find that the BPS condition on the supersymmetry parameters simply reads as
\begin{align}
    \label{BPSds}
        \overbar{\xi}^{i\alpha} = \xi^{i\alpha}~.
\end{align}
The same relation holds with the index $\alpha$ lowered.

\subsection*{The free $\cN=2$ hyper-multiplet}
This multiplet is formed by two complex scalars, $q$ and $\widetilde{q}$, two chiral fermions $\psi^a_\alpha$ and two anti-chiral fermions $\overbar{\psi}^{a\dot\alpha}$. Organizing the scalars in a $(2\times 2)$ matrix $Q^{ia}$ as in (\ref{Qin}), namely
\begin{align}
    \big(Q^{ia}\big)=\begin{pmatrix}
        {q}^*&\widetilde{q}~\\
        -\widetilde{q}^{\,*}&q
    \end{pmatrix} ~,
    \label{Qinapp}
\end{align}
the supersymmetry transformations are (suppressing spinor indices)
\begin{subequations}
    \begin{align}
        &\delta Q^{ia}=-\sqrt{2}\,\xi^{i}\psi^{a}-\sqrt{2}\,\bar{\xi}^i\overbar{\psi}^a~,\\
        &\delta\psi^a=\ii\,\sqrt{2}\,\bar{\xi}^i\overbar{\sigma}^\mu
        \partial_\mu Q^{ja}\epsilon_{ji}~,\\
        &\delta \overbar{\psi}^a=-\ii\,\sqrt{2}\,{\xi}^i{\sigma}^\mu
        \partial_\mu Q^{ja}\epsilon_{ji}~.
    \end{align}
    \label{SUSYhyper}%
\end{subequations}
They close the algebra (\ref{SUSYalgebra}) on shell.

\subsection*{The free $\cN=2$ current multiplet}
This multiplet consists of three scalars $\Phi^{ij}=\Phi^{ji}$, such that
$(\Phi^{ij})^*=\epsilon_{ik}\,\epsilon_{j\ell}\,\Phi^{k\ell}$, two chiral fermions $X_{i\alpha}$,
two anti-chiral fermions $\overbar{X}_i^{\,\dot\alpha}$, two real scalars $P$ and $\overbar{P}$, and one conserved current $j_\mu$. These fields can be regarded as a subset of the $\cN=4$ stress tensor multiplet. Their supersymmetry transformations are (suppressing spinor indices) \cite{Dolan:2001tt}
\begin{subequations}
    \begin{align}
        &\delta \Phi^{ij}=\frac{1}{2}\,\xi^{i}\epsilon^{jk}X_{k}-\frac{1}{2}\,\bar\xi^i\epsilon^{jk}\overbar{X}_k+ (i\leftrightarrow j)~,\label{deltaPhiij}\\
        &\delta X_{i}=\xi^{j}\epsilon_{ji}\,P +2 \,\ii \, \overbar{\xi}^{j}\,\overbar{\sigma}^\mu \partial_{\mu} \Phi^{k\ell} \,\epsilon_{kj}\,\epsilon_{\ell i} + \overbar{\xi}^j\epsilon_{ji}\,\overbar{\sigma}^\mu j_{\mu} ~,\\[1mm]
        &\delta\overbar{X}_i=\overbar{\xi}^j\epsilon_{ji} \,\overbar{P} +2 \,\ii\,
        \xi^{j}\, \sigma^\mu \partial_\mu\Phi^{k\ell}\, \epsilon_{k j}\,\epsilon_{\ell i} + \xi^{j}
        \epsilon_{ji}\,\sigma^\mu j_\mu~,\\[1mm]
        &\delta P=- 2 \,\ii\,\overbar{\xi}^i\,\overbar{\sigma}^\mu \partial_{\mu} X_{i}~,\quad
        \delta\overbar{P}= 2\, \ii \,\xi^{i}\,\sigma^\mu \partial_{\mu}\overbar{X}_{i}~,\\[1mm]
        &\delta j_{\mu}=-\ii \,\xi^{i} \,\sigma_{\mu\nu}\,\partial^{\nu}X_{i} +
        \ii\,\overbar{\xi}^{i}\,\overbar{\sigma}_{\mu\nu}\,\partial^{\nu}\overbar{X}_{i}~.
    \end{align}
    \label{SUSYcurrent}%
\end{subequations}
These transformation close the $\cN=2$ algebra (\ref{SUSYalgebra}) if $\partial^\mu j_\mu=0$. One can verify that these supersymmetry transformations follow from those in (\ref{SUSYhyper}) upon using the
definitions given in (\ref{flavorhyper}). In this calculation one also sees that the on-shell conditions
on the hyper-multiplet components imply the conservation of $j_\mu$.

\section{Stereographic projection}
\label{appendix:projection}
A 4-sphere $S^4$ can be conveniently described as embedded in $\mathbb{R}^5$. Let the coordinates of $\mathbb{R}^5$ be $(\eta^\mu,\eta^5)$ with $\mu=1,\cdots,4$. Then, a 4-sphere with unit radius is described by the following equation
\begin{equation}
    \sum_{\mu=1}^4(\eta^\mu)^2+(\eta^5)^2=1~.
\end{equation}
In these coordinates, the North and South poles are the points:
\begin{align}
    \mathrm{N}:= (0,0,0,0,+1)\quad\mbox{and}\quad
    \mathrm{S}:= (0,0,0,0,-1)~.
    \label{poles}
\end{align}
We now conformally map the $S^4$ to $\mathbb{R}^4$ with coordinates $x^\mu$ by means of the following stereographic projection
\begin{align}
    \eta^\mu=\frac{2x^\mu}{1+x^2}~,\quad
    \eta^5=\frac{1-x^2}{1+x^2}
    \label{stereographic}
\end{align}
where $x^2=\sum_{\mu}(x^\mu)^2$.
It is easy to see that under such a map we have
\begin{equation}
    ds^2=\sum_{\mu=1}^4(d\eta^\mu)^2+(d\eta^5)^2=\tilde{\Omega}(x)^{-2}\,\sum_{\mu=1}^4 (dx^\mu)^2
    \label{metric}
\end{equation}
where the conformal factor is
\begin{equation}
    \tilde{\Omega}(x)=\frac{1+x^2}{2}~.
    \label{omegatilde}
\end{equation}
Under this projection the North and South poles are mapped respectively to the origin and to the point at
infinity in $\mathbb{R}^4$.

\begin{figure}[ht]
    \centering
    \includegraphics[scale=0.55]{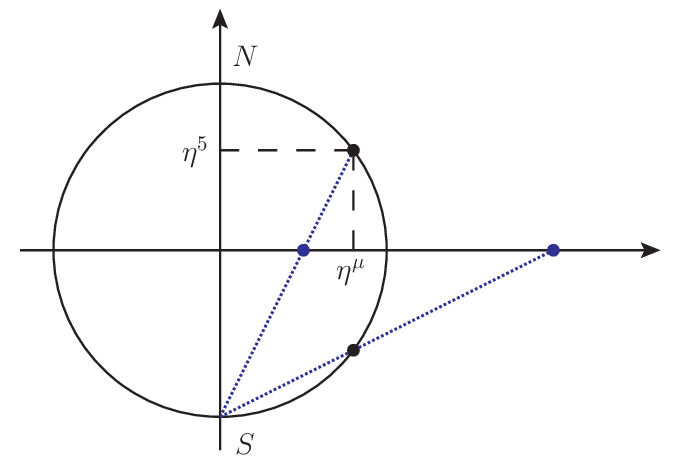}
    \caption{Stereographic projection of a vertical great circle of $S^4$ onto a straight line in $\mathbb{R}^4$.}
    \label{fig:vertical_loop}
\end{figure}

A great circle $\mathcal{C}$ of $S^4$ passing through the North and South poles (\ref{poles}) can be parametrized as 
\begin{equation}
    (0,0,0,\sin\tau,\cos\tau)\quad\mbox{with}\quad \tau\in[-\pi,+\pi]~.
    \label{verticalloop}
\end{equation}
Under the stereographic projection (\ref{stereographic}), it is mapped to a straight line in $\mathbb{R}^4$ parametrized by
\begin{equation}
    (0,0,0,t)\quad\mbox{with}\quad t=\tan\Big(\frac{\tau}{2}\Big)~.
    \label{line}
\end{equation}
This is shown in Figure~\ref{fig:vertical_loop}.
\section{Ward identities}
\label{appendix:ward}
\subsection*{The $\mathcal{N}=2$ current multiplet in presence of a Wilson line}
\label{subsec:currentWilson}
Let us consider a $\mathcal{N}=2$ current multiplet in presence of a Wilson line. Its supersymmetry transformations can be obtained from those listed in (\ref{SUSYcurrent}) by applying the effective
rules (\ref{rulesds}) to reduce the rotational symmetry to the one preserved by the defect and by imposing the identification
(\ref{BPS}) to guarantee the invariance of Wilson line. To write
these transformations in a simpler form, it is convenient to introduce the following combinations\,%
\footnote{In the original notation we would have $Y_{i\alpha}=X_{i\alpha}+\ii\,\overbar{X}_i^{\,\dot\alpha}$ and $Z_{i\alpha}=X_{i\alpha}-\ii\,\overbar{X}_i^{\,\dot\alpha}$.}
    \begin{align}
     Y_{i\alpha} &= X_{i\alpha} - \bar X_{i\alpha}~,~~
     Z_{i\alpha} = X_{i\alpha} + \bar X_{i\alpha}~,~~
      S  = P -\overbar{P} - 2\, j^4~,~~
      T = P + \overbar{P}~.
     \label{defST}
\end{align}
Using these combinations and understanding the contractions on spinor indices, one obtains
\begin{subequations}
\begin{align}
\label{susyphi}
\delta\Phi^{ij} & = -\frac{1}{2}\xi^{i}\,Y^{j}+(i\leftrightarrow j)~,\\
\label{susyY}
\delta Y^i & = S\,\xi^i - 4 \,\ii\, \partial_4\Phi^{ij}\,\xi_j~,\\
\label{susyS}
\delta S & = 2\,\ii\,\xi^i \,\partial_4 Y_i~,\\
\label{susyZ}
\delta Z^i & = T\,\xi^i - 2\,\ii\, \vec j\cdot\vec\sigma\,\xi^i 
+ 4\,\vec\partial\Phi^{ij}\cdot\vec\sigma\,\xi_j~,\\
\label{susyT}
\delta T & = 2\,\ii\,\xi^i\,\partial_4Z_i + 2\xi^i\,\vec\sigma\cdot\vec\partial Y_i~,\\
\label{susyj}
\delta \vec j & = - \xi^i\, \vec\sigma\, \partial_4 Z_i
+ \xi^i \,\vec\sigma\times \vec\partial Y_i~.
\end{align}
\label{SUSYcurrentBPS}%
\end{subequations}
One can easily verify that, on all fields in the multiplet, these supersymmetries square to the translations along the Wilson line as in (\ref{SUSYalgebraBPS}). 
It is also interesting to observe that the fields $\Phi^{ij}$, $Y_{i\alpha}$ and $S$ form a sub-multiplet on which the supersymmetry algebra closes.

\subsection*{Constraints from $R$ and flavour symmetry}
The 2-point correlators of fields belonging to the current multiplet that are allowed by statistics (namely boson/boson and fermion/fermion ones) are restricted in form by $R$-symmetry. For example, since $\Phi^{ij}$ is symmetric in $i$ and $j$ and $\epsilon^{ij}$ is the only $R$-invariant 2-index tensor at our disposal, we have
\begin{align}
    \label{formcorr1}
       \big\langle \Phi^{ij}(x_1)\,S(x_2)\big\rangle_W & = \big\langle \Phi^{ij}(x_1)\,T(x_2)\big\rangle_W 
       = \big\langle \Phi^{ij}(x_1)\,\vec j(x_2)\big\rangle_W = 0
\end{align}
and
\begin{align}       
    \label{formcorr2}
        \big\langle \Phi^{ij}(x_1)\,\Phi^{k\ell}(x_2)\big\rangle_W 
        & = (\epsilon^{ik}\,\epsilon^{j\ell} + \epsilon^{i\ell}\,\epsilon^{jk}) A(x_1,x_2)
\end{align}
where $A(x_1,x_2)$ a symmetric function of its arguments.
Similarly, for the fermion/fermion correlators we have
\begin{subequations}
    \begin{align}
        \big\langle Y^i_\alpha(x_1)\,Y^j_\beta(x_2)\big\rangle_W 
        & = \epsilon^{ij}\big(\epsilon_{\alpha\beta}\, b(x_1,x_2) + \vec\tau_{\alpha\beta}\cdot \vec b(x_1,x_2)\big)~,\\[1mm]
        \big\langle Z^i_\alpha(x_1)\,Y^j_\beta(x_2)\big\rangle_W 
        & = \epsilon^{ij}\big(\epsilon_{\alpha\beta}\, c(x_1,x_2) + \vec\tau_{\alpha\beta}\cdot \vec c(x_1,x_2)\big)~,\\[1mm]
        \big\langle Z^i_\alpha(x_1)\,Z^j_\beta(x_2)\big\rangle_W 
        & = \epsilon^{ij}\big(\epsilon_{\alpha\beta}\, d(x_1,x_2) + \vec\tau_{\alpha\beta}\cdot \vec d(x_1,x_2)\big)~.
\end{align}
 \label{formcorr3}%
\end{subequations}
where $\Vec{\tau}_{\alpha\beta}\equiv\vec{\sigma}_{\alpha}^{~\gamma}\,\epsilon_{\gamma\beta}$ is symmetric in $\alpha$ and $\beta$. The functions $b(x_1,x_2)$ and $d(x_1,x_2)$ are anti-symmetric in their arguments, while $\vec b(x_1,x_2)$ and $\vec d(x_1,x_2)$ are symmetric. We further define
\begin{align}
    \big\langle j^\mu(x_1)\,j^\nu(x_2)\big\rangle_W&=C^{\mu\nu}(x_1,x_2)~,
    \label{jmujnu}\\[1mm]
        \big\langle T(x_1)\, T(x_2)\big\rangle_W &= H(x_1,x_2)~.
            \label{lstcorr}
\end{align}

Also the flavour symmetry imposes restrictions on the form of the 2-point functions. To see this, we recall that the operators $P$ and $\overbar{P}$ contain terms with opposite non-zero flavour charges, as explicitly shown in (\ref{PPbar}). Thus, they can have non-vanishing correlation functions among themselves\,%
\footnote{Notice that the correlator $\big\langle P(x_1)\, P(x_2)\big\rangle_W$ is not zero through its $\tr(\psi^1\psi^1)\tr(\psi^2\psi^2)$ term. Indeed, there is a bulk Yukawa coupling of the form $\overbar{\varphi}\overbar{\psi}^1\overbar{\psi}^2$ which can turn a $\psi^1\psi^2$ pair into a $\overbar{\varphi}$ that can be adsorbed by the Wilson line. A mirror reasoning holds for the $\big\langle \overbar{P}(x_1)\,\overbar{P}(x_2)\big\rangle_W$ correlator.}. Instead, 
their correlators with the current $j^\mu$, which is neutral, vanish. This fact entails that 
\begin{subequations}
    \begin{align}
        \big\langle  j^\mu(x_1)\,T(x_2)\big\rangle_W & = 0~,\\[1mm]
        \big\langle j^\mu(x_1)\, S(x_2)\big\rangle_W & = -2 \big\langle j^\mu(x_1)\, j^4(x_2)\big\rangle_W
        = -2\, C^{\mu4}(x_1,x_2)~,  \label{vecjT}\\[1mm]
        \big\langle S(x_1)\,S(x_2)\big\rangle_W & = \big\langle Q(x_1)\,Q(x_2)\big\rangle_W + 4\, \big\langle j^4(x_1)\,j^4(x_2)\big\rangle_W  = L(x_1,x_2) +4\, C^{44}(x_1,x_2)~.
        \label{SSform}
\end{align}
\end{subequations}
For convenience, in the last line we have defined $Q=P-\overbar{P}$ and set $\big\langle Q(x_1)\,Q(x_2)\big\rangle_W=L(x_1,x_2)$.
It is quite natural to assume that $\big\langle P(x_1)\,P(x_2)\big\rangle_W = \big\langle\overbar{P}(x_1)\,\overbar{P}(x_2)\big\rangle_W $ and that $\big\langle P(x_1\,\overbar P(x_2)\big\rangle_W = \big\langle \overbar{P}(x_1)\, P(x_2)\big\rangle_W$. From this, it follows that
\begin{align}
    \label{TS}
        \big\langle T(x_1)\, S(x_2)\big\rangle_W=\big\langle T(x_1)\, Q(x_2)\big\rangle_W = \big\langle (P(x_1)+\overbar{P}(x_1))\, (P(x_2)-\overbar{P}(x_2))\big\rangle_W = 0~.  
\end{align}
Notice that even if do not make this assumption \emph{a priori}, the consistency of the supersymmetry Ward identities forces \emph{a posteriori} the correlator between $T$ and $S$ to vanish, as we will shortly see.

\subsection*{Constraints from supersymmetry} 
By considering the supersymmetry variation of boson/fermion correlators, which are obviously vanishing, one can obtain relations among the above 2-point functions, which further constrain their form. 
For example, starting from $\big\langle \Phi^{ij}(x_1)\,Y^k_\alpha(x_2)\big\rangle_W=0$, we get
\begin{align}
        0 &= \big\langle \delta\Phi^{ij}(x_1)\,Y^k_\alpha(x_2)\big\rangle_W \,\epsilon_{kj}
        + \big\langle \Phi^{ij}(x_1)\,\delta Y^k_\alpha(x_2)\big\rangle_W \,\epsilon_{kj} \label{phiY1}\\[1mm]
        &= \Big[-\frac 12 \big\langle Y^j_\beta(x_1) \,Y^k_\alpha(x_2)\big\rangle_W \,\xi^{i\beta} - \frac 12 \big\langle Y^i_\beta(x_1)\, Y^k_\alpha(x_2)\big\rangle_W \,\xi^{j\beta} - 4\,\ii \big\langle \Phi^{ij}(x_1)\,\partial_4\Phi^{k\ell}(x_2)\big\rangle_W\, \xi_{\ell,\alpha}\Big]
	\epsilon_{kj}~.\notag
\end{align}
Taking into account the constraints (\ref{formcorr1}), (\ref{formcorr2}) and (\ref{formcorr3}), and contracting the $\epsilon$-symbols, we obtain
\begin{align}
    \label{phiY2a}
        0 = b(x_1,x_2)\, \xi^i_\alpha + \vec b(x_1,x_2)\cdot \vec\sigma_{\alpha}^{~\beta}\, \xi^i_\beta - 8 \,\ii\, \partial^{(2)}_4A(x_1,x_2)\,\xi^i_\alpha~,
\end{align}
where with $\partial^{(a)}_\mu$ we denote the derivative with respect to $x_a^\mu$. This relation implies
\begin{align}
    \label{btoA}
    \vec b(x_1,x_2) = \vec 0~,~~~
        b(x_1,x_2) = 8\,\ii\, \partial^{(2)}_4 A(x_1,x_2)
         = -8\,\ii\, \partial^{(1)}_4 A(x_1,x_2)~,
\end{align}
where in the last step we have taken into account that $b(x_1,x_2)$ is anti-symmetric, and thus $A(x_1,x_2)$ can depend on the $4$th components only through the combination $x_{12}^4=x_1^4-x_2^4$, in agreement with translational invariance along the direction of the line defect.

Consider now $\vev{S(x_1)\,Y^i_\alpha(x_2)}_W = 0$. From its supersymmetry variation we obtain
\begin{align}
    0 &= \big\langle \delta S(x_1)\,Y^i_\alpha(x_2)\big\rangle_W + \big\langle S(x_1)\,\delta Y^i_\alpha(x_2)\big\rangle_W\notag\\[1mm]
    & = -2\,\ii \,\big\langle \partial^{(1)}_4 Y^j_\beta(x_1)\, Y^i_\alpha(x_2)\big\rangle_W\,\xi^\beta_j + \big\langle S(x_1)\, S(x_2)\big\rangle_W \,\xi^i_\alpha
        \label{varSY}
\end{align}
where in the second line we have inserted the $R$-symmetry constraint (\ref{formcorr1}).
Using (\ref{formcorr3}) and (\ref{SSform}) and exploiting (\ref{btoA}), this reduces to
\begin{align}
    \label{SY3}
        L(x_1,x_2) + 4 \,C^{44}(x_1,x_2) =  2\,\ii \, \partial^{(1)}_4 b(x_1,x_2)
        = -16\, \partial^{(1)}_4 \partial^{(2)}_4 A(x_1,x_2)~.
\end{align}

Starting from $\big\langle Z^k_\alpha(x_1)\,\Phi^{ij}(x_2)\big\rangle_W=0$, we find
\begin{align}
        0 & = \big\langle \delta Z^k_\alpha(x_1)\,\Phi^{ij}(x_2)\big\rangle_W\,\epsilon_{kj}
        + \big\langle Z^k_\alpha(x_1)\,\delta\Phi^{ij}(x_2)\big\rangle_W \,\epsilon_{kj}\label{varZphi}\\[1mm]
        &= \Big[
         4 \,\big\langle \vec\partial^{(1)}\Phi^{k\ell}(x_1)\,\Phi^{ij}(x_2)\big\rangle_W \cdot \vec\sigma_{\alpha}^{~\beta}\, \xi_{\ell\beta}
        +\frac 12 \big\langle Z^k_\alpha(x_1)\,Y^j_\beta(x_2)\big\rangle_W \,\xi^{i\beta} + \frac 12 \big\langle Z^k_\alpha(x_1)\,Y^i_\beta(x_2)\big\rangle_W\, \xi^{j\beta} \Big]\,\epsilon_{kj}\notag
\end{align}
where as before we have retained only the correlators that are not zero by global symmetry considerations.
Using (\ref{formcorr1}), (\ref{formcorr2}) and (\ref{formcorr3}) and contracting the $\epsilon$-symbols, we obtain
\begin{align}
    \label{phiY2}
        0 = - 8\, \vec\partial^{(1)}A(x_1,x_2)\cdot
        \vec\sigma_{\alpha}^{~\beta}\,\xi^{i}_\beta + c(x_1,x_2)\, \xi^i_\alpha + \vec c(x_1,x_2)\cdot\vec\sigma_{\alpha}^{~\beta}\, \xi^i_\beta~,
\end{align}
implying
\begin{align}
    \label{ctoA}
        c(x_1,x_2) = 0~,~~~
        \vec c(x_1,x_2) = 8\, \vec\partial^{(1)}A(x_1,x_2)~.
\end{align}

From the variation of $\big\langle Z^i_\alpha(x_1)\,S(x_2)\big\rangle_W=0$, it follows that
\begin{align}
    \label{ZSvar1}
        0 & = \big\langle \delta Z^i_\alpha(x_1)\, S(x_2)\big\rangle_W + 
        \big\langle Z^i_\alpha(x_1)\, \delta S(x_2)\big\rangle_W
        \notag\\[1mm]
        & = \big\langle T(x_1)\,S(x_2)\big\rangle_W \,\xi^i_\alpha - 2 \,\ii \big\langle \vec j(x_1)\, S(x_2)\big\rangle_W \cdot \vec\sigma_{\alpha}^{~\beta}\, \xi^i_\beta
        - 2 \,\ii \,\big\langle Z^i_\alpha(x_1)\, \partial^{(2)}_4 Y^j_\beta(x_2)\big\rangle_W \,\xi^{j\beta}~. 
\end{align}
Using (\ref{formcorr3}), after simple algebra this becomes
\begin{align}
    \label{ZSvar2}
        0 & = \big\langle T(x_1)\,S(x_2)\big\rangle_W \,\xi^i_\alpha - 2\, \ii\,\Big[ \big\langle j^m(x_1)\, S(x_2)\big\rangle_W 
        - \partial^{(2)}_4 c^m(x_1,x_2)\Big] 
        \cdot (\sigma^m)_{\alpha}^{~\beta}\, \xi^i_\beta~,
\end{align}
which requires
\begin{align}
    \label{ZSvar3}
        \big\langle T(x_1)\,S(x_2)\big\rangle_W = 0~,~~~
        \big\langle j^m(x_1)\, S(x_2)\big\rangle_W =  \partial^{(2)}_4 c^m(x_1,x_2)
        = 8\, \partial^{(1)}_m\,\partial^{(2)}_4 A(x_1,x_2)
\end{align}
where in the last step we inserted the relation (\ref{ctoA}). As anticipated, the vanishing of the correlator between $T$ and $S$, which we imposed as a consequence of the flavor symmetry, has been now obtained as a constraint imposed by the supersymmetry Ward identities.

If we consider the variation of $\big\langle T(x_1)\,Y^i_\alpha(x_2)\big\rangle_W=0$, we do not get any new information. 
Indeed, one can show that
\begin{align}
    \label{TYvar}
        0 & = \big\langle \delta T(x_1)\, Y^i_\alpha(x_2)\big\rangle_W + \big\langle T(x_1)\, \delta Y^i_\alpha(x_2)\big\rangle_W=\Big[2\,\ii\, \partial^{(1)}_4\vec c(x_1,x_2) + 2\, \vec\partial^{(1)} b(x_1,x_2)\Big]\cdot \vec\sigma_{\alpha}^{~\beta}\,\xi^i_\beta ~.
\end{align}
with the quantity in square brackets identically vanishing if we use the results 
(\ref{btoA}) and (\ref{ctoA}). 

Next we consider the variation of $\big\langle Z^i_\alpha(x_1)\, T(x_2)\big\rangle_W=0$. Proceeding as in previous cases, we find
\begin{align}
        0 & = \big\langle \delta Z^i_\alpha(x_1)\,T(x_2)\big\rangle_W +\big\langle Z^i_\alpha(x_1)\, \delta T(x_2)\big\rangle_W \label{ZTvar1}\\[1mm]
        &= \big\langle T(x_1)\,T(x_2)\big\rangle_W\,\xi^i_\alpha
        + 2\,\ii\, \xi_j^\beta\,\big\langle Z^i_\alpha(x_1)\,\partial^{(2)}_4 Z^j_\beta(x_2)\big\rangle_W+ 2 \,\xi_j^\beta \,\vec{\sigma}_\beta^{~\gamma} \big\langle Z^i_\alpha(x_1)\, \vec{\partial}^{(2)}Y^j_\gamma(x_2)\big\rangle_W\notag~.
\end{align}
Inserting the form (\ref{formcorr3}) and (\ref{lstcorr}) of the correlators and taking into account that $c(x_1,x_2)=0$, after some 
algebraic manipulations this relation becomes
\begin{align}
    \label{ZTvar3}
        0 & = \Big[H(x_1,x_2) + 2\,\ii\, \partial^{(2)}_4d(x_1,x_2) - 2 \,\vec\partial^{(2)} \cdot \vec c(x_1,x_2) \Big]\,\xi^i_\alpha \notag\\[1mm]
        &\quad+ 2 \,\ii\Big[\partial^{(2)}_4\vec d(x_1,x_2) + \vec\partial^{(2)}\times\vec c(x_1,x_2)\Big]\cdot
        \vec\sigma_{\alpha}^{~\beta}\,\xi^i_\beta~.
\end{align}
The two square brackets must separately vanish, leading to 
\begin{subequations}
    \begin{align}
         H(x_1,x_2) + 2\,\ii\, \partial^{(2)}_4d(x_1,x_2) - 16\, \vec\partial^{(2)} \cdot \vec\partial^{(1)} A(x_1,x_2) &= 0~,  \label{ZTvar41}\\
        \partial^{(2)}_4\vec d(x_1,x_2) + 8\, \vec\partial^{(2)} \times \vec\partial^{(1)} A(x_1,x_2) &= 0  \label{ZTvar42}
\end{align}
\label{dvecd}%
\end{subequations}
where we have used the expression of $\vec{c}(x_1,x_2)$ given in
(\ref{ctoA}).

Starting from $\big\langle Z^i_\alpha(x_1)\,j^n(x_2)\big\rangle_W=0$, we get
\begin{align}
        0 & = \big\langle \delta Z^i_\alpha(x_1)\, j^n(x_2)\big\rangle_W + \big\langle Z^i_\alpha(x_1)\,\delta j^n(x_2)\big\rangle_W \label{Zjvar}\notag\\[1mm]
        & = - 2\, \ii \,\big\langle j^m(x_1)\, j^n(x_2)\big\rangle_W \,(\sigma^m)_{\alpha}^{~\beta}\,\xi_\beta - \xi^\beta_j\, (\sigma^n)_\beta^{~\gamma} 
        \big\langle Z^i_\alpha(x_1)\, \partial^{(2)}_4Z^j_\gamma(x_2)\big\rangle_W \\[1mm]
        & \quad + \epsilon^{nmp}\, \xi^\beta_j \,(\sigma^m)_{\beta}^{~\gamma}\, \big\langle Z^i_\alpha(x_1)\, \partial^{(2)}_p Y^j_\gamma(x_2)\big\rangle_W~.\notag
\end{align}
Using the parametrizations (\ref{formcorr3}) and (\ref{jmujnu}) of the correlators, after some algebra we can recast this relation in the form
\begin{align}
    \label{Zjvar2}
        0 & = - 2\, \ii\, C^{mn}(x_1,x_2)\, (\sigma^m)_\alpha^{~\beta}\,\xi^i_\beta + \partial^{(2)}_4d(x_1,x_2)\,(\sigma^n)_{\alpha}^{~\beta}\,\xi^i_\beta
        + \partial^{(2)}_4d^n(x_1,x_2)\,\xi^i_\alpha \notag\\ 
        & \quad- \ii \,\epsilon^{nmp} \partial^{(2)}_4 d^m(x_1,x_2)\, (\sigma^p)_\alpha^{~\beta}\,\xi^i_\beta 
        + \epsilon^{nmp} \partial^{(2)}_mc^p(x_1,x_2)\, \xi^i_\alpha \\
        & \quad+ \ii\, \partial^{(2)}_m c^m(x_1,x_2)\, (\sigma^n)_\alpha^{~\beta}\,\xi^i_\beta
        -\ii\,\partial^{(2)}_mc^n(x_1,x_2)\, (\sigma^m)_\alpha^{~\beta}\,\xi^i_\beta~.\notag 
\end{align}
The condition for the vanishing of the coefficient of $\xi^i_\alpha$ is
\begin{align}
    \label{Zjvar3}
        0 & = \partial^{(2)}_4 \vec d(X_1,X_2) + \vec\partial^{(2)}\times \vec c(x_1,x_2)
\end{align}
which, taking into account (\ref{ctoA}), coincides with (\ref{ZTvar42}). Thus, this condition does not add any new constraint. Instead, by imposing the vanishing of the coefficient of $(\sigma^m)_\alpha^{~\beta}\,\xi^i_\beta$, we extract the following new relation
\begin{align}
    \label{Zjvar4}
        2\,C^{mn}(x_1,x_2) + \big[\ii\,\partial^{(2)}_4 d(x_1,x_2) - \vec\partial^{(2)}\cdot \vec c(x_1,x_2)\big]\,\delta^{mn} 
        + \epsilon^{mnp} \partial^{(2)}_4 d^p(x_1,x_2) + \partial^{(2)}_m c^n(x_1,x_2) = 0~,
\end{align}
which, using (\ref{ctoA}), can be rewritten as
\begin{align}
    \label{Zjvar5}
         2\, C^{mn}(x_1,x_2) + \big[\ii\,\partial^{(2)}_4 d(x_1,x_2) 
         - 8\, \vec\partial^{(2)}\cdot \vec\partial^{(1)} A(x_1,x_2)\big]\, \delta^{mn}
         + 8\, \partial^{(1)}_m\partial^{(2)}_nA(x_1,x_2) = 0~.
\end{align}
In the coefficient of $\delta^{mn}$ we recognize the same combination that appears in (\ref{ZTvar41}), so that in the end we have
\begin{align}
    \label{Zjvar6}
        C^{mn}(x_1,x_2) = \frac 14 \,H(x_1,x_2)\,\delta^{mn} - 4\, \partial^{(1)}_m\partial^{(2)}_n A(x_1,x_2)~.
\end{align}

The last boson/fermion correlator to consider is $\big\langle j^m(x_1)\, Y^i_\alpha(x_2)\big\rangle_W=0$. However, one can show that its supersymmetry variation yields the following two conditions
\begin{align}
        \partial^{(1)}_4c^m(x_1,x_2) + \big\langle j^m(x_1)\,S(x_2)\big\rangle_W = 0~,~~~
        -\ii\, \partial^{(1)}_4c^n(x_1,x_2) - \partial^{(1)}_n b(x_1,x_2) = 0~,
\end{align}
which are identically satisfied taking into account the earlier results (\ref{ZSvar3}) and (\ref{btoA}). Thus, no new constraints are obtained from this correlator.

We summarize the identities we have found in Tab.~\ref{table:susyWI}.
\begin{table}[ht]
\begin{center}
\begin{tabular}{ |c| }
\hline
$\phantom{\bigg|}\big\langle \Phi^{ij}(x_1)\,\Phi^{k\ell}(x_2)\big\rangle_W=(\epsilon^{ik}\,\epsilon^{j\ell}+\epsilon^{i\ell}\,\epsilon^{jk}\big)\,A(x_1,x_2)$ \\\hline
$\phantom{\bigg|}\big\langle Y^i_\alpha(x_1)\,Y^j_\beta(x_2)\big\rangle_W=8\,\ii\,\epsilon^{ij}\,\epsilon_{\alpha\beta}\,\partial^{(2)}_4A(x_1,x_2)$ \\\hline
$\phantom{\bigg|}\big\langle Z^i_\alpha(x_1)\,Y^j_\beta(x_2)\big\rangle_W=8\,\epsilon^{ij}\,\vec{\tau}_{\alpha\beta}\cdot \vec{\partial}^{(1)}_4A(x_1,x_2)$ \\\hline
$\phantom{\bigg|}\big\langle Z^i_\alpha(x_1)\,Z^j_\beta(x_2)\big\rangle_W= \epsilon^{ij}\,\epsilon_{\alpha\beta}\, d(x_1,x_2) + 
\epsilon^{ij}\,\vec\tau_{\alpha\beta}\cdot \vec d(x_1,x_2)$ \\\hline
$\phantom{\bigg|}\big\langle S(x_1)\,S(x_2)\big\rangle_W=-16\, \partial^{(1)}_4 \partial^{(2)}_4 A(x_1,x_2)$ \\\hline
$\phantom{\bigg|}\big\langle j^m(x_1)\,S(x_2)\big\rangle_W=8\, \partial^{(1)}_m\,\partial^{(2)}_4 A(x_1,x_2)$ \\\hline
$\phantom{\bigg|}\big\langle T(x_1)\,T(x_2)\big\rangle_W=H(x_1,x_2)$ \\\hline
$\phantom{\bigg|}\big\langle j^m(x_1)\,j^n(x_2)\big\rangle_W=\frac 14 \,H(x_1,x_2)\,\delta^{mn} - 4\, \partial^{(1)}_m\partial^{(2)}_n A(x_1,x_2)~~
$ \\\hline
\end{tabular}
\end{center}
\caption{The non-vanishing 2-point functions of the component fields of the current multiplet. The functions $d(x_1,x_2)$ and
$\vec{d}(x_1,x_2)$ appearing in the $Z$-$Z$ correlator are related to the functions $A(x_1,x_2)$ and $H(x_1,x_2)$ as indicated in (\ref{dvecd}).}
\label{table:susyWI}
\end{table}

\subsection*{Constraints from defect CFT}
We now show that by imposing consistency with the rules of the
defect CFT \cite{Billo:2016cpy} one obtains further constraints on the form of the 2-point functions derived in the previous subsection. 

In our derivation we follow \cite{Herzog:2020bqw} and introduce the two invariant cross-ratios preserved by the conformal symmetry in presence of a Wilson line given by\,%
\footnote{In \cite{Herzog:2020bqw} the invariants $u$ and $v$ are called, respectively, $\xi_1$ and $\xi_2$. Since in our discussion also the supersymmetry parameters have been called $\xi_i$ we have preferred to use a different name for the cross-ratios.}
\begin{equation}
    u=\frac{(x_1-x_2)^2}{4\,r_1 r_2}~,~~~
    v=\frac{\vec{x}_1\cdot\Vec{x}_2}{r_1r_2}~.
    \label{uv}
\end{equation}
These invariants are related to $\xi$ and $\eta$ used in the main text and defined in (\ref{eq:CrossRatios}) as follows
\begin{equation}
u=\frac{\xi-\eta}{2}~,~~~v=\eta~.
    \label{uvvsxieta}
\end{equation}

We recall that the top-components $\Phi^{ij}$ of the current multiplet are conformal fields of dimension 2. Thus, their 2-point function in presence of the Wilson line has the form (\ref{eq:2pt_W}) with $\Delta_1=\Delta_2=2$. This implies that
\begin{equation}
    A(x_1,x_2)=\frac{a}{r_1^2r_2^2}
    \label{auv}
\end{equation}
where $a$ is a function of the cross-ratios $u$ and $v$. Also the 2-point function $\big\langle T(x_1)\,T(x_2)\big\rangle$ has a similar form, but with $\Delta_1=\Delta_2=3$. Thus,
\begin{equation}
    H(x_1,x_2)=\frac{h}{r_1^3r_2^3}
    \label{huv}
\end{equation}
where $h$ is a function of $u$ and $v$. Our goal is to establish a relation between the functions $h$ and $h$ by exploiting the Ward identities previously obtained. To do so, we use the relation among the
current/current correlator, the derivatives of $A$ and the function $H$ as displayed in the last row of Tab.~\ref{table:susyWI}. The crucial ingredient for this calculation is the form of the current/current correlator in a defect CFT. This has been derived in full generality in \cite{Lauria:2018klo,Herzog:2020bqw}. Adapting eq.~(3.4) of \cite{Herzog:2020bqw} to our present conventions and recalling that $j^\mu$ has conformal dimension 3, we have
\begin{subequations}
    \begin{align}
    \big\langle j^m(x_1)\,j^n(x_2)\big\rangle_W&=\frac{1}{r_1^3r_2^3}\bigg[\delta^{mn}\Big(\frac{f_4+f_5}{64u^3}\Big)
    +\frac{x_1^mx_2^n}{r_1r_2}\Big(\frac{f_4}{128u^4}+\frac{(1+4u^2)f_1}{256u^5}+\frac{f_2+2f_3}{64u^3}\Big)\notag\\[1mm]
    &\qquad~~~~+\frac{x_1^nx_2^m}{r_1r_2}\Big(\frac{f_4}{128u^4}-\frac{f_5}{64u^3 v}+\frac{f_1}{256u^5}+
\frac{f_2}{64u^3 v^2}-\frac{f_3}{64u^4 v}\Big)\notag\\[1mm]
    &\qquad~~~~+\frac{x_1^mx_1^n+x_2^mx_2^n}{r_1 r_2}\Big(\frac{f_3}{128u^4  v}-\frac{f_1}{256u^5}-\frac{f_4}{128u^4}\Big)\notag\\[1mm]
    &\qquad~~~~+\Big(\frac{x_1^mx_2^n}{r_1^2}+\frac{x_1^mx_2^n}{r_2^2}\Big)\Big(\!\!-\frac{f_1+f_3}{128u^4}\Big)\notag\\[1mm]
    &\qquad~~~~+\Big(\frac{x_1^mx_1^n}{r_1^2}+\frac{x_2^mx_2^n}{r_2^2}\Big)\Big(\frac{f_1}{128u^4}-\frac{f_2}{64u^3v}+\frac{(v-2u)f_3}{128u^4v}\Big)\bigg]~,\label{jmjn}\\[2mm]
     \big\langle j^m(x_1)\,j^4(x_2)\big\rangle_W&=-\frac{x_{12}^4}{r_1^3r_2^3}\bigg[\frac{x_1^m}{r_1 r_2}\Big(\frac{f_1}{256u^5}+\frac{f_4}{128u^4}\Big)+\frac{x_2^m}{r_1 r_2}\Big(\frac{f_3}{128u^4v}-\frac{f_1}{256u^5}-\frac{f_4}{128u^4}\Big)\notag\\[1mm]
&\qquad~~~~~~+ \frac{x_1^m}{r_1^2}\Big(\!\!-\frac{f_1+f_3}{128u^4}\Big)\bigg]~,\label{jmj4}\\[2mm]
 \big\langle j^4(x_1)\,j^4(x_2)\big\rangle_W&=\frac{1}{r_1^3\,r_2^3}
 \bigg[\frac{f_4}{64u^3}-\frac{(x_{12}^4)^2}{r_1r_2}\Big(\frac{f_1}{256\,u^5}+\frac{f_4}{128\,u^4}\Big)\bigg]~,\label{j4j4}
\end{align}
\label{jj}%
\end{subequations}
where $f_1,\ldots, f_5$ are five functions of the invariants $u$ and $v$. 

We are now in the position of deriving the constraints imposed by the Ward identities.
From
\begin{equation}
    \big\langle j^m(x_1)\,j^n(x_2)\big\rangle_W -\frac{1}{4}H(x_1,x_2)\,\delta^{mn}+4\partial_m^{(1)}\partial_n^{(2)}A(x_1,x_2)=0~,
\end{equation}
using (\ref{auv}), (\ref{huv}) and (\ref{jmjn}), computing the derivatives of $A(x_1,x_2)$ and imposing the vanishing of the six independent coordinate structures, we obtain a linear system of six equations that allows 
us to write the functions $f_1,\ldots,f_5$ and $h$ in terms of the derivatives of $a$.
The solution of this system turns out to be
\begin{subequations}
    \begin{align}
    f_1&=-256u^4\big[u\,\partial_u^2a+3\partial_ua-4v(2a+u\,\partial_ua+v\,\partial_va)\big]~,\\
    f_2&=-256u^3v\big[v\,\partial_v^2a+3\partial_va-4u(2a+u\,\partial_ua+v\,\partial_va)\big]~,\\
    f_3&=-256u^4v\big[\partial^2_{u,v}a+4(2a+u\,\partial_ua+v\,\partial_va)\big]~,\\
    f_4&=+128u^3\big[3\partial_ua-4(2u+v)(2a+u\,\partial_ua+v\,\partial_va)\big]~,\\
    f_5&=-256u^3\big[3\partial_va-2(2u+v)(2a+u\,\partial_ua+v\,\partial_va)\big]~,
    \end{align}
    \label{f1f5}%
\end{subequations}
and
\begin{equation}
    h=16(\partial_ua-2\partial_va)~.
    \label{his}
\end{equation}
Inserting this last result in (\ref{huv}) we then obtain
\begin{equation}
    \big\langle T(x_1)\,T(x_2)\big\rangle_W=\frac{16\partial_ua-32\partial_va}{r_1^3 r_2^3}~.
    \label{TTfin}
\end{equation}
On the other hand, using the Ward identity (\ref{ZSvar3}) together with the fact that $\langle j^m(x_1)\,Q(x_2) \rangle=0$ because of the flavor symmetry constraints\,\footnote{
We observe that even without imposing the vanishing of this correlator, one would nevertheless conclude that it has to vanish for consistency. Indeed, as shown in \cite{Herzog:2020bqw} the 2-point function of a current and a scalar operator in a defect CFT contains structures that appear neither in the $j^m$-$j^4$ correlator nor in the double derivatives of the scalar function $A$ with respect to $x_1^m$ and $x_2^4$. Thus, the only way to impose the Ward identity constraint is to require that the correlator of $j^m$ and $Q$ vanishes, just as required by flavor symmetry.}, we obtain the following relation
\begin{align}
    \big\langle j^m(x_1)\,j^4(x_2)\big\rangle_W +4\,\partial_{(1)}^m\partial_{(2)}^4A(x_1,x_2)=0~.
    \label{jmj4new}
\end{align}
Exploiting (\ref{jmj4}) and proceeding as before, we obtain from (\ref{jmj4new}) a linear system of three equations for the functions $f_1$, $f_3$ and $f_4$ that appear in the $j^m$-$j^4$ correlator. The solution of this system is
\begin{subequations}
    \begin{align}
    f_1&=-256u^4\big[u\,\partial_u^2a+3\partial_ua\big]~,\\
    f_3&=-256u^4v \,\partial^2_{u,v}a~,\\
    f_4&=+384u^3 \partial_ua
    \end{align}
    \label{f1f3f4}%
\end{subequations}
which is compatible with (\ref{f1f5}) provided
\begin{equation}
    2a+u\partial_ua+v\partial_va=0~.
    \label{omogeneity}
\end{equation}
As a further consistency check, we consider the Ward identity 
(\ref{SY3}), namely
\begin{equation}
    \big\langle Q(x_1)\,Q(x_2)\big\rangle_W+4\big\langle j^4(x_1)\,j^4(x_2)\big\rangle_W+16\,\partial_4^{(1)}\partial_4^{(2)}A(x_1,x_2)=0~.
    \label{QQj4j4}
\end{equation}
Using (\ref{j4j4}) and the expression for $f_1$ and $f_4$ that we have found before together with the condition (\ref{omogeneity}), it is easy to deduce from (\ref{QQj4j4}) that
\begin{align}
    \big\langle Q(x_1)\,Q(x_2)\big\rangle_W&=\frac{1}{r_1^3r_2^3}
    \bigg[\Big(\!\!-\frac{f_4}{16u^3}+8\partial_ua\Big)+\frac{(x_{12}^4)^2}{r_1r_2}\Big(\frac{f_1}{64u^5}+\frac{f_4}{32u^4}+4\partial_u^2a\Big)\bigg]
    =-\frac{16\partial_ua}{r_1^3r_2^3}~.
    \label{QQfin}
\end{align}
This has exactly the form expected for a 2-point function of scalar operators of dimension 3 in a defect CFT.

We can now rewrite our results on the scalar correlators in terms of the cross-ratios $\xi$ and $\eta$ used in the main text. From (\ref{uvvsxieta}) we easily see that $\partial_u=2\partial_\xi$ and $\partial_v=\partial_\xi+\partial_\eta$, so that (\ref{TTfin}) and (\ref{QQfin}) simply become
\begin{equation}
\big\langle T(x_1)\,T(x_2)\big\rangle_W=-\frac{32\,\partial_\eta a}{r_1^3r_2^3}~~~\mbox{and}~~~
    \big\langle Q(x_1)\,Q(x_2)\big\rangle_W=-\frac{32\,\partial_\xi a}{r_1^3r_2^3}~.
    \label{TTQQfin}
\end{equation}
In terms of these cross-ratios, the homogeneity equation (\ref{omogeneity}) reads $2a+\xi\partial_\xi a+\partial_\eta a=0$.
We conclude by mentioning that we have also performed the above analysis using the parametrization of the current/current correlators in the defect CFT given in \cite{Lauria:2018klo} with the embedding formalism. The results obtained with this method are identical to the ones we have presented.

\printbibliography

\end{document}